%% file: main.tex
\documentclass[twocolumn]{aastex631}

\usepackage{natbib}
\usepackage{amsmath,empheq}
\usepackage{graphicx}
\usepackage{textcomp}
\usepackage{mathtools}
\usepackage{gensymb}    
\usepackage{xspace}
\usepackage{hyperref}
\usepackage{longtable}
\usepackage[encapsulated]{CJK}
\usepackage{ucs}
\usepackage[utf8x]{inputenc}

\newcommand{\galbase}{\texttt{galbase}\xspace}
\newcommand{\scanam}{\texttt{Scanamorphos}\xspace}
\newcommand{\zzmgs}{$z$0MGS\xspace}
\newcommand{\dustbff}{\texttt{DustBFF}\xspace}
\newcommand{\ngint}{877\xspace}
\newcommand{\ngres}{819\xspace}

\newcommand{\umin}{$U_{\rm min}$\xspace}
\newcommand{\sigd}{$\Sigma_{\rm d}$\xspace}

\newcommand{\qpah}{$q_{\rm PAH}$\xspace}

\newcommand{\omgs}{$\Omega_*$\xspace}
\newcommand{\ubar}{$\overline{U}$\xspace}

\newcommand{\mdust}{M$_{\rm d}$\xspace}
\newcommand{\stm}{M$_\star$\xspace}
\newcommand{\dms}{$\Delta{\rm MS}$\xspace}
\newcommand{\dmsr}{$\Delta{\rm MS}_{\rm R}$\xspace}
\newcommand{\sigstm}{$\Sigma_\star$\xspace}
\newcommand{\sigsfr}{$\Sigma_{\rm SFR}$\xspace}

\newcommand{\ssfrr}{sSFR$_{\rm R}$\xspace}
\newcommand{\rtf}{r$_{25}$\xspace}
\newcommand{\gasfr}{$\langle \Sigma_{\rm SFR} \rangle$\xspace}
\newcommand{\gamstar}{$\langle \Sigma_{*} \rangle$\xspace}
\newcommand{\gamd}{$\langle \Sigma_{\rm d} \rangle$\xspace}
\newcommand{\msolpcsq}{M$_\odot$~pc$^{-2}$\xspace}
\newcommand{\msolkpcsqyr}{M$_\odot$~kpc$^{-2}$~yr$^{-1}$\xspace}

\newcommand{\herschel}{{\em Herschel}\xspace}
\newcommand{\logt}{log$_{10}$}
\newcommand{\kms}{km~s$^{-1}$}
\newcommand{\hii}{\ion{H}{2}\xspace}
\newcommand{\msun}{M$_{\odot}$\xspace}

\shorttitle{Resolved dust properties of 800 galaxies}
\shortauthors{Chastenet, Sandstrom et al.}

\begin{document}


\title{The Resolved Behavior of Dust Mass, Polycyclic Aromatic Hydrocarbon Fraction, \\ and Radiation Field in $\sim$~800 Nearby Galaxies}

\email{jeremy.chastenet@ugent.be} 

\author[0000-0002-5235-5589]{J\'er\'emy~Chastenet}
\affiliation{Sterrenkundig Observatorium, Universiteit Gent, Krijgslaan 281 S9, B-9000 Gent, Belgium}

\author[0000-0002-4378-8534]{Karin~Sandstrom}
\affiliation{Department of Astronomy \& Astrophysics, University of California, San Diego, 9500 Gilman Drive, La Jolla, CA 92093, USA}

\author[0000-0002-2545-1700]{Adam~K.~Leroy}
\affiliation{Department of Astronomy, The Ohio State University, 4055 McPherson Laboratory, 140 West 18th Ave, Columbus, OH 43210, USA}

\author[0000-0001-6118-2985]{Caroline~Bot}
\affiliation{Observatoire Astronomique de Strasbourg, Universit\'e de Strasbourg, UMR 7550, 11 rue de l’Universit\'e, F-67000 Strasbourg, France}

\author[0000-0003-2551-7148]{I-Da~Chiang
\begin{CJK*}{UTF8}{bkai}(江宜達)\end{CJK*}}
\affiliation{Institute of Astronomy and Astrophysics, Academia Sinica, No. 1, Sec. 4, Roosevelt Road, Taipei 106216, Taiwan}

\author[0000-0001-8241-7704]{Ryan~Chown}
\affiliation{Department of Astronomy, The Ohio State University, 140 West 18th Avenue, Columbus, OH 43210, USA}

\author[0000-0001-5340-6774]{Karl~D.~Gordon}
\affiliation{Space Telescope Science Institute, 3700 San Martin Drive, Baltimore, MD 21218, USA}

\author[0000-0001-9605-780X]{Eric~W.~Koch}
\affiliation{Center for Astrophysics $\mid$ Harvard \& Smithsonian, 60 Garden Street, Cambridge, MA 02138, USA}

\author{H\'el\`ene~Roussel}
\affiliation{Sorbonne Universit\'e, Institut d’Astrophysique de Paris, CNRS (UMR7095), 75014 Paris, France}

\author[0000-0002-9183-8102]{Jessica~Sutter}
\affiliation{Whitman College, 345 Boyer Avenue, Walla Walla, WA 99362, USA}

\author[0000-0002-0012-2142]{Thomas~G.~Williams}
\affiliation{Subdepartment of Astrophysics, Department of Physics, University of Oxford, Keble Road, Oxford OX1 3RH, UK}



\begin{abstract}
We present resolved 3.6--250 $\micron$ dust spectral energy distribution (SED) fitting for $\sim 800$ nearby galaxies. We measure the distribution of radiation field intensities heating the dust, the dust mass surface density (\sigd), and the fraction of dust in the form of polycyclic aromatic hydrocarbons (PAHs; \qpah). We find that the average interstellar radiation field (\ubar) is correlated both with stellar mass surface density (\sigstm) and star formation rate surface density (\sigsfr), while more intense radiation fields are only correlated with \sigsfr. We show that \qpah\ is a steeply decreasing function of \sigsfr, likely reflecting PAH destruction in \hii regions. Galaxy integrated \qpah\ is strongly, negatively correlated with specific star formation rate (sSFR) and offset from the star-forming ``main sequence'' (\dms), suggesting that both metallicity and star formation intensity play a role in setting the global \qpah. We also find a nearly constant M$_{\rm d}$/M$_*$ ratio for galaxies on the main sequence, with a lower ratio for more quiescent galaxies, likely due to their lower gas fractions. From these results, we construct prescriptions to estimate the radiation field distribution in both integrated and resolved galaxies. We test these prescriptions by comparing our predicted \ubar\ to results of SED fitting for stacked ``main sequence'' galaxies at $0<z<4$ from \citet{Bethermin2015} and find sSFR is an accurate predictor of \ubar\ even at these high redshifts.  Finally, we describe the public delivery of matched-resolution WISE and {\em Herschel} maps along with the resolved dust SED fitting results through the InfraRed Science Archive (IRSA). 
\end{abstract}



\section{Introduction}
Interstellar dust is fundamentally important to the physics of the interstellar medium (ISM). By absorbing photons over a wide range of wavelengths and converting their energy to grain heating (which is then re-radiated primarily in the infrared) or gas heating (via the ejection of photo-electrons), dust couples starlight to the ISM. The re-radiated starlight also serves as a tracer of star formation observable across a wide redshift range \citep{KennicuttEvans12KSReview,Tacconi2020}. The amount, composition, and size of grains govern their absorption and scattering efficiencies, infrared emissivity, photoelectric yields, and other key properties \citep[e.g.,][]{Draine2003,Galliano2018,Hensley2023,Ysard2024}.  It is therefore of great importance for a wide range of astrophysical topics to understand what sets the abundance and properties of dust and the radiation field illuminating it.

Infrared (IR) emission is an important tool to study dust because it represents the most accessible direct tracer of dust and it can be observed across a wide range of galaxy types and redshifts. Extinction and elemental depletions provide crucial complementary constraints on grain size distributions and composition \citep[for a recent compilation of such constraints see][and references therein]{Hensley21}. However, in galaxies outside the Local Group, the necessary measurements to study depletion and extinction can only be done in small samples of sight-lines with sufficiently bright background point sources \citep[e.g., bright stars in relatively nearby galaxies, or background quasars that sample dust in their foreground;][]{Jenkins2009, Peroux2020, Roman-Duval2022}. Measurements of dust attenuation can also provide insights into dust content and properties, but complexities related to geometry and scattering make such studies challenging \citep[for a review, see][]{SalimNarayaban20}. By contrast, dust emission is detectable from the near-IR through millimeter and has been observed extensively by ground, stratospheric, and space-based facilities. 

In particular, the \herschel Space Observatory \citep[][]{Pilbratt10}, over the course of its mission, observed far-IR dust emission in a large sample of resolved, nearby galaxies, at 70, 100, and 160 \micron\ using the Photodetector Array Camera and Spectrometer \citep[PACS;][]{Poglitsch2010}  and at 250, 350, and 500~$\mu$m with the Spectral and Photometric Imaging REceiver \citep[SPIRE;][]{Griffin2010}. \herschel observations will remain the best-quality (in sensitivity and angular resolution) far-IR dataset until the next generation far-IR facility comes into service. The set of nearby galaxies observed with \herschel PACS and SPIRE are therefore a critical resource for understanding dust properties, and are the focus of this study.

The basic features of the IR spectral energy distribution (SED) of dust emission are set by 1) the dust mass surface density (\sigd), 2) the dust composition (including the fraction of dust in the form of polycyclic aromatic hydrocarbons, PAHs; \citeauthor{ATB89} \citeyear{ATB89}; see also reviews from \citeauthor{Tielens08} \citeyear{Tielens08}, \citeauthor{Li20PAHs} \citeyear{Li20PAHs}), and 3) the intensity (and potentially the spectrum) of the radiation field illuminating the dust\footnote{We assume optically thin emission in the mid- to far-IR throughout this study. In the nearby galaxies we study, significant optical depth in the IR is rare. Attenuation can be an issue for ultra-luminous infrared galaxies (ULIRGs).}. The intensity of the radiation field sets the steady-state temperature of the grains in equilibrium with the radiation field. This temperature can be inferred from the modified blackbody-like emission that the equilibrium grains produce (peaking typically between 100--200 \micron). Given the equilibrium temperature and knowledge of the dust grain optical properties, the dust mass surface density, \sigd, can be inferred from the intensity of the equilibrium IR emission. Small grains are not in equilibrium with the radiation field, but rather are stochastically heated to high temperatures by the absorption of single UV photons \citep{Sellgren1984,Draine1985,Draine2001}. A key component of the small grain population are PAHs, which produce bright vibrational emission bands in the mid-IR after absorbing a UV/optical photon.  Given the intensity of the illuminating radiation field constrained by the equilibrium grains, the mid-IR emission allows one to infer the fraction of the dust mass in the form of PAHs \citep{Draine07} and the distribution of radiation field intensities.

Within the beam of typical extragalactic IR observations (hundreds of pc to kpc scales), a range of radiation field intensities will inevitably be present \citep{Helou1986,Dale01,dale2002}.  It is therefore necessary to make some assumptions about the sub-resolution radiation field distribution in modeling the IR SED.  This is particularly important because of the very steep dependence of dust luminosity on temperature, which can lead to biased temperatures \citep[e.g.,][]{Utomo19}.  
One common approach, based on work by \citet{Dale01} and further expanded upon by \citet{Draine07} involves a power-law distribution of radiation field intensities, with a delta-function at a minimum intensity (\umin) meant to represent the overall diffuse ISM interstellar radiation field. The assumed distribution of radiation field intensities is a critical aspect of interpreting the dust SED. In models with a single radiation field, the width of the equilibrium dust peak and mid-IR emission can be interpreted as indicators of changing dust properties, particularly variations in very small grain abundance or dust composition \citep[see][Figure 3 for a clear visualization]{Galliano2018}.  

A variety of studies have performed IR SED fitting in nearby galaxies, both resolved \citep{Munoz-Mateos2009, Aniano12, Gordon14, Hunt2015, Casasola2017, Aniano20, Chastenet21, Clark2021, Abdorrouf2022, Casasola2022, Clark2023} and unresolved  \citep{Boselli2010, remy-ruyer2014, Davies2017, Clark2018, Galliano2021, Dale2023}, using various dust SED models. These studies have revealed the existence of scaling relationships between dust mass, radiation field, and PAH fraction with galaxy properties and examined the correlations of the parameters with the local galactic environment.  Key trends that have been identified include the scaling of dust mass with stellar mass and metallicity \citep{daCunha2010, Cortese2012, Calura2017, Galliano2021}; PAH fraction with stellar mass, metallicity, and radiation field intensity \citep{Nersesian19, Aniano20, Galliano2022}; and radiation field parameters with star formation rate and stellar mass \citep{Nersesian19}.  In general, samples where resolved dust SED fitting have been performed are often small, focusing on $\sim$~tens of targets \citep[e.g., ][]{Aniano20,Casasola2022}. The samples studied with integrated SED fits are typically much larger, up to $\sim900$ galaxies as in the Dustpedia\footnote{\url{http://dustpedia.astro.noa.gr}} compilation \citep{Davies2017, Bianchi2018, Clark2018, Clark2019, Davies2019, Nersesian2019, Casasola2020, Galliano2021}.

In the following work, we extend resolved analysis to $\sim 800$ nearby galaxies with observations from \herschel. This study focuses on deriving resolved dust properties in a large sample of nearby galaxies, and the relationship between dust and stellar parameters, such as stellar mass and star formation rate surface density. Covering a wide range of both dust and stellar properties is essential to fully appreciate the interplay between ISM properties and galaxy evolution. The archive of the \textit{Herschel} space mission is a goldmine for a resolved study of dust property variations in the $z=0$ universe.

The paper is organized as follows. In Section~\ref{SecSample} we describe the definition of the \textit{Herschel} archival sample, which includes all nearby galaxies with PACS or SPIRE observations.  
In Section~\ref{SecDataProcessing} we describe the reduction of the archival Herschel data, and the additional data processing we perform.
We present the details of the fitting procedure in Section~\ref{SecFitting}, and a description of the results in Section~\ref{SecResults}.
In Section~\ref{SecDustProperties}, we discuss the new insights into the properties of dust and radiation fields across the local galaxy population.
Our conclusions are summarized in Section~\ref{SecConclusions} and we describe our data delivery to the NASA Infrared Science Archive (IRSA) (doi: \href{https://www.ipac.caltech.edu/doi/10.26131/IRSA581}{{\tt 10.26131/IRSA581}}\footnote{\url{https://irsa.ipac.caltech.edu/data/Herschel/z0MGS_Dust/overview.html}}) in Appendix~\ref{AppDelivery}.

\section{Galaxy Sample}
\label{SecSample}
Our galaxy sample is defined as the overlap between the nearby galaxies in the HyperLeda\footnote{\url{http://leda.univ-lyon1.fr/}} database \citep{HyperLeda} and those that are found to be covered by either a PACS or SPIRE photometric observation in the Herschel Science Archive\footnote{\url{http://archives.esac.esa.int/hsa/whsa/}}. We begin with the full sample of galaxies at $cz < 5\,000$ \kms\ in HyperLeda. To create this selection we query the database for all galaxies (i.e., object type `G') with measured heliocentric radial velocities $<5\,000$ \kms. This returned a sample of 33,322 objects. We then queried the Herschel Science Archive with that list to find all observations where the polygon defining the field of view of the observation intersects with a circle of radius of 1$'$ centered at the galaxy's central coordinates. This search was done for all PACS and SPIRE photometric observing modes, aside from parallel modes.  The query results in 1\,245 matches for PACS observations and 1\,729 for SPIRE (note that some of these are repeated observations of the same galaxies). We also included three very nearby galaxies observed in PACS-SPIRE parallel mode: M31, M33, and IC~342. When a galaxy is observed multiple times, we use all observations (see Section~\ref{sec:multipleobs}). Some of our targets lie at the edge of \textit{Herschel} scans and are eliminated from the sample after visual inspection of each image. 

Some nearby galaxies were also observed as part of large-area surveys for studying galaxy evolution at higher redshift, as in the large map of the Virgo Cluster \citep[the Herschel Virgo Cluster Survey, HeViCs;][]{Davies10hevics}.  Due to the computational resources required to reduce large area maps, we do not include those in our reprocessing of archival data.  In the case of HeViCs, we extract galaxies from data products already delivered to the Herschel Archive, as described below. Combining the individual pointed observations in PACS or SPIRE; HeViCs; and M31, M33, and IC~342; our initial sample includes 1\,580 unique galaxies with some Herschel observations.

Although we will only use galaxies with both PACS and SPIRE (as well as WISE) observations in the dust SED fitting described below, we have reduced all galaxies with any PACS or SPIRE maps that are not in a deep field observation and provide those in our data delivery (described in Appendix~\ref{AppDelivery}). 

In the following, we use distances and orientation parameters from the HyperLeda database collated in the \texttt{galbase} code\footnote{\url{https://github.com/akleroy/galbase}}, updated in some cases to reflect detailed studies of small samples (e.g., using rotation curve fitting). The list of adopted positions, distances, and orientation parameters for all targets in our sample can be found in the delivery table and a list of the contents of the catalog is provided in Table~\ref{tab:sample} \citep[adopted parameters are mostly the same as in][]{Leroy19}. For the SED fitting analysis, we narrow our sample to the galaxies with WISE mid-IR observations from the $z=0$ Multiwavelength Galaxy Synthesis project \citep[z0MGS;][]{Leroy19}, as the mid-IR bands are necessary to perform robust fits using a physical dust model.
This leads to a dataset of \ngint galaxies.

\input{tab_delivery}

\begin{figure}
    \centering
    \includegraphics[width=0.5\textwidth]{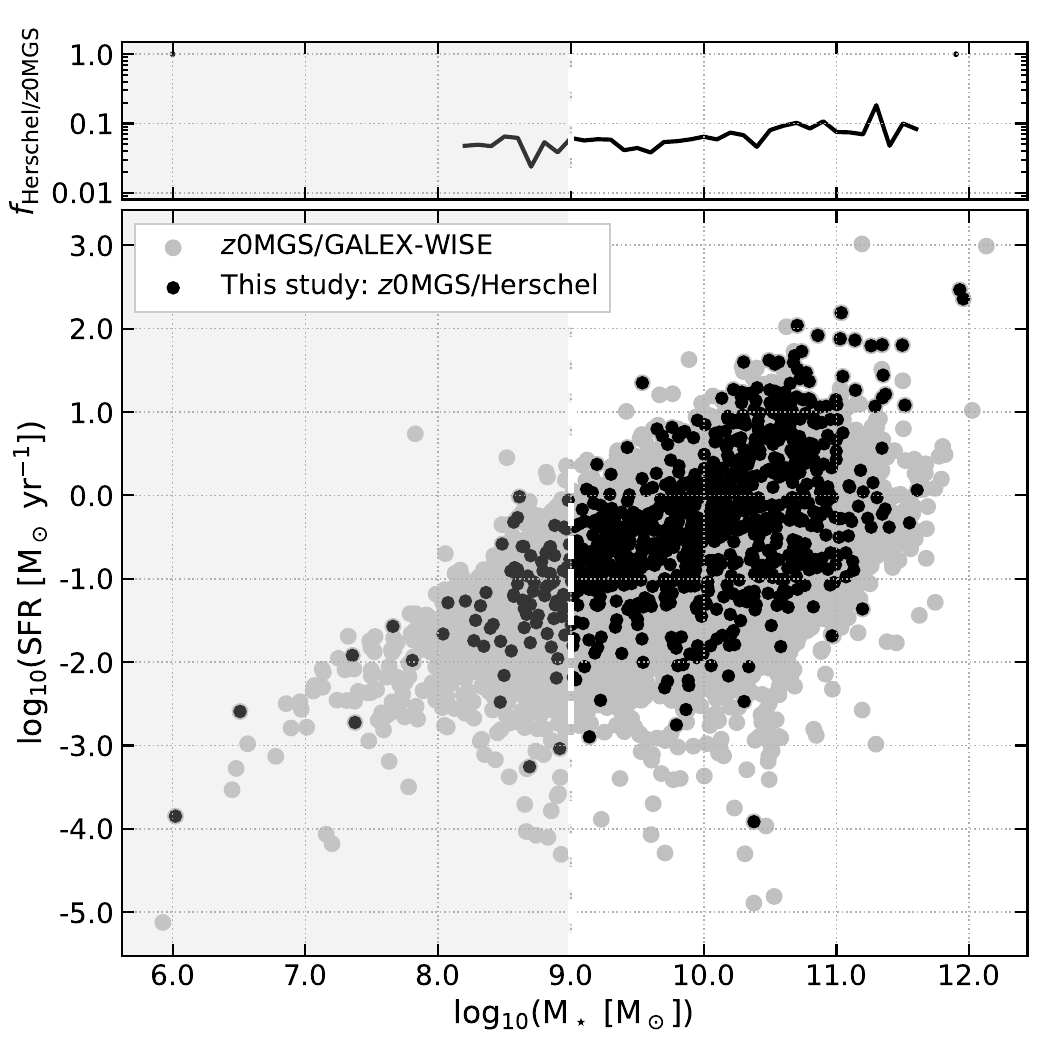}
    \caption{Stellar mass vs star formation rate for the \zzmgs sample \citep[grey points,][]{Leroy19} highlighting the galaxies with WISE \textit{and} \textit{Herschel} maps used in this study as black points. Above $10^9$~\msun, the \zzmgs sample is approximately complete for galaxies within 50 Mpc. The top panel shows the relative number of galaxies in our sample compared to the \zzmgs sample in bins of \stm. We find that $\sim 10$\% of the \zzmgs galaxies have {\em Herschel} coverage and are included in our sample. That fraction does not change substantially with stellar mass.}
    \label{FigSampleProp}
\end{figure}

In Figure~\ref{FigSampleProp} we show our galaxy sample overlaid on the full sample from \zzmgs in \stm--SFR space. The \zzmgs sample is approximately complete for galaxies above $10^9$~\msun within 50~Mpc.
The top panel of Figure~\ref{FigSampleProp} shows the fraction of galaxies in our \textit{Herschel} sample compared to the \zzmgs sample, in bins of stellar mass. That fraction is relatively constant above $10^9$~\msun, where the reference sample is complete.  We see that $\sim$10\% of the \zzmgs sample above $10^9$~\msun is included in our analysis. The SFR--M$_\star$ space is filled relatively homogeneously within the completeness bounds \zzmgs sample. This suggests that the selection of galaxies with \textit{Herschel} observations is representative of the population of nearby galaxies, allowing us to draw conclusions about the typical behavior of radiation fields and dust properties for the $z\sim0$ galaxy population. 

\section{Data Reduction \& Analysis}
\label{SecDataProcessing}
After the initial sample definition from the query to the \textit{Herschel} Science Archive, we downloaded the Level~0 data for all bands and all targets and reduced them with the \texttt{HIPE} and \scanam pipelines (see Section~\ref{SecDataFIR}). This initial sample includes 1580 galaxies. Note that the the SPIRE bands at 350 and 500~$\mu$m are not used to derive dust properties (see Section~\ref{SecFitting}), but we have reduced those bands with the same processing steps described below and they are included in our public data delivery.
In the following sections, we describe in more detail the data reduction and processing that was applied to all galaxies.  We then describe the further data processing needed for the SED fitting analysis, including the incorporation of the \zzmgs WISE data.

\subsection{Far-IR: Herschel}
\label{SecDataFIR}
\subsubsection{Initial processing}
The PACS and SPIRE queries to the \textit{Herschel} Science Archive were done independently, and we require in both cases that the target was observed in any of the photometry modes. From the queries, we extract the target name and observation IDs for each source. We processed the parallel-mode observations for M31, M33, and IC~342 in a similar manner, simply changing the relevant keyword in \scanam to reduce them as such. 
We note that for very extended galaxies on the sky, additional processing would be needed to recover very diffuse, extended emission, following \citet{Clark2023}. At present, our reduction does not account for any filtered large scale emission.

We use \texttt{HIPE} \citep[Herschel Interactive Processing Environment, version 15;][]{Wieprecht09, Dowell10, Ott10} and the  \scanam  suite of routines \citep[version 25, released August 2016;][]{Roussel13} to process scans from L0 to L2 products (we use the PACS calibration tree v.~77, and the SPIRE calibration tree v.~14.3). This first step takes the raw data, formats it into images, and removes the instrumental noise, as described in \citet[][]{Roussel13}. It returns FITS files containing the formatted image (in Jy/pix for PACS, and Jy/beam for SPIRE) and associated uncertainty, a weight map, and a drift map.
The weight map reflects the number of scans per pixel, and informs our assessment of the instrumental noise. We use these maps to weight pixels with respect to this number of scans.

\subsubsection{Targets with more than one observation}\label{sec:multipleobs}
A few galaxies in our sample were targeted more than once throughout \herschel operations. When this is the case, we manually combine all scans targeting the galaxy into the same formatting framework before running the pipeline.
In \scanam, this is done by appending scan filenames within the same call to the formatting function.
This is valid when observation are done with similar modes. 
Observations in ``mini-maps'' mode were included with the appropriate keywords. We do not include observations that target supernovae to avoid adding in a time variable source.

\subsubsection{Extracting galaxies from HeViCS}
\label{SecHevics}
HeViCS \citep[][]{Davies10hevics, irsa70}\footnote{\url{http://www.hevics.org/}} is a large program that mapped the Virgo galaxy cluster using the PACS~100 \micron\ through the SPIRE~500 \micron\ bands by observing four contiguous large fields.  The reduced HeViCS data have been delivered to HSA as a high-level science product\footnote{\url{https://irsa.ipac.caltech.edu/data/Herschel/HeVICS/overview.html}}.  
HeViCS includes twelve galaxies that have also been observed with targeted observations in other programs. 
We compare our reprocessing of these 12 galaxies to the HeViCS high-level products by extracting fluxes within \rtf. We find good agreement between the fluxes, which confirms that a full reprocessing of the large HeViCS mosaics is not necessary. As such, we use the products from HeViCS in line with those produced by our reprocessing. For galaxies covered by HeViCS that also have dedicated observations we use the dedicated observations. We extract a total of 65 galaxies from HeViCS.

To extract galaxies from HeViCS, we select a 5~\rtf area in each band around the central coordinates of each galaxy.
In comparing to WISE~4, we found a small astrometric offset between the extracted \textit{Herschel} and the WISE images for most of these galaxies. We used WISE as a standard and correct the extracted HeViCS images. This is done by finding the translation (we assume the misalignment is translation only) that maximizes the correlation between the reference image (WISE) and the newly extracted HeViCS cutout.  We then update the headers of the images to reflect the adjusted coordinates. After these two steps, we process the HeViCS maps the same way as our other data, including background removal, convolution to a common resolution and regridding.

\subsection{Mid-IR: WISE}
We combine our far-IR data with the WISE \citep[][]{Wright10} images compiled and analyzed in the the z0MGS project \citep{Leroy19}. WISE mapped the near- and mid-IR sky at $\lambda\sim 3.4$, 4.6, 12, and 22~\micron~(hereafter WISE~1, WISE~2, WISE~3, and WISE~4). We refer to \citet[][]{Leroy19} for the detailed description of the catalog creation and image processing. We use their data products at $15''$ resolution, which we will then convolve to match the resolution of the SPIRE 250 \micron\ observations.

In some specific cases, the WISE cut-outs provided by \citet[][]{Leroy19} are much smaller than the \herschel image and significantly limit the available pixels to measure the background covariance matrix (Section~\ref{SecFittingProcedure}). For these galaxies, we extract larger cutouts from the WISE tiles to ensure proper measurement of the background. This does not affect any other aspect of the fitting.

\subsection{Unit Conversion and Background Removal}
\label{SecBkgRemoval}
The output from the initial reduction for PACS and SPIRE observations are in Jy/pixel and Jy/beam, respectively. We convert PACS data to MJy/sr using the pixel size. The SPIRE data are converted to MJy/sr and corrected from point source to extended source calibration using the ${\rm K_{PtoE}}$ factors in the SPIRE Handbook v.17\footnote{\url{http://herschel.esac.esa.int/Docs/SPIRE/spire_handbook.pdf}}.

Far-IR observations are subject to several foreground/background contributions, including MW cirrus, zodiacal light, instrumental offsets, etc. To remove these, we perform a background subtraction with a similar approach to that in \citet{Leroy19} and \citet{Utomo19} using a 2D plane.
For each target, we create a mask that is used to exclude galaxy emission from the background plane fit. We mask:
\vspace{-0.2cm}
\begin{itemize}
\setlength\itemsep{-0.2em}
    \item all pixels within a radius $r=A \times {\rm r_{25}}$ from the galaxy center. The default value is $A=1.5$. We require a minimum of 50 pixels to be masked within that radius. If this criteria is not met, A is increased until this inner mask has this minimum required area. Alternatively, A is decreased if the numbers of unmasked pixels (i.e., background pixels) is less than 10\% of the total number of pixels in the image. No bounds are applied to vary A, but its final values range between 1 and 1.6, for the whole sample;
    \item other known galaxies in the field of view, identified from HyperLeda, are masked with their own coordinates, and $A \times r_{25}$ as described above;
    \item from the weight map given by Level~1 \scanam products, we mask pixels that are below the median value of that weight map. This ensures that we use only the most well-covered pixels to estimate the background;
    \item finally, we perform an iterative clipping, masking any pixel above the median $+3\times$ the standard deviation of the background pixels. This is done several times, recomputing median and standard deviation, until the fractional difference between two calculated values of the median, $m_{i}^{\rm bkg}$, $m_{i+1}^{\rm bkg}$, is less than 1\%.
\end{itemize}
The final set of unmasked pixels are fit with a 2D plane representing the background. This plane is then subtracted from the image. In our data product delivery we provide the background-subtracted image, the final background mask, and store the coefficients describing the subtracted plane in the headers.

\subsection{Convolution and re-gridding}
We perform convolutions to put all mid- and far-IR images at matched resolution prior to our pixel-by-pixel SED modeling.  First, all WISE and PACS data are convolved to the circularized SPIRE~250 point spread function (PSF; ${\rm FWHM} \sim 18''$). This provides the highest angular resolution for the dust SED fitting results, but retains the characteristics of the SPIRE 250 PSF which has some non-Gaussianity. For convolution we use the circularized kernels from \citet{Aniano2011} for PACS-to-SPIRE convolutions. We also convolve the native resolution SPIRE 250 to a circularized SPIRE 250 PSF, following \citet{Aniano2011}. For WISE, we use a kernel to convolve a $15''$-Gaussian to circularized SPIRE~250 also from \citet{Aniano12}. For each target, we re-project all convolved images to the pixel grid of the SPIRE~250 image. This grid has pixel size $4.5''$ and so oversamples the SPIRE beam with $\sim 18''$ FWHM, as has been done in other works \citep[e.g.,][]{Paradis2023}. Another possible approach seen in the literature is to set the pixel size to that of the PSF. In that case, pixels can be considered independent, carrying little correlation between them, and treated as unique data points \citep[e.g.,][]{Gordon14, Viaene2014, Vutisalchavakul2014, Casasola2015, Saikia2020, Casasola2022}.
In our case, the analysis is carried on a large sample that we  bin, and extract medians, and the oversampling will have a limited effect.

\subsection{Background Covariance Matrix Masks}
\label{SecMasks}
As described in Section~\ref{SecFitting}, our SED fit makes use of a covariance matrix describing the background at each wavelength included in the fit.  To determine this for each galaxy, we need to select which pixels to attribute to the background only, i.e., which are free of any target galaxy or bright foreground star emission (effectively, faint stars and background galaxies are included in the covariance matrix as confusion noise). These pixels are used to measure the background covariance matrix (described in Section~\ref{SecFittingProcedure}). We construct this mask using the background-subtracted images after convolution and regridding. To identify background pixels in these final images, we follow the same steps used to estimate the background in Section~\ref{SecBkgRemoval} adopting the already determined $A$ coefficient from that step. The bright pixels identified in any band during iterative masking are excluded from the covariance estimation, as are the regions affected by bright stars in the mid-IR bands in the \zzmgs delivery. 

\subsection{Integrated Galaxy Photometry}
\label{SecIntegratedPhotometry}
We use aperture photometry to derive the integrated flux within \rtf\ for each band and each target. We also estimate an associated uncertainty by scaling down the value of the pixel-by-pixel noise measured from a signal-free region outside of \rtf by the square root of the number of resolution elements in \rtf. 
These measurements are included in our delivered data products, summarized in Table~\ref{tab:sample}.

At this stage, we trim a subset of galaxies from the sample based on the signal-to-noise of their integrated photometry. 
If the integrated flux of any available PACS or SPIRE bands is below $1\sigma$, we do not perform an integrated SED fitting on that target (and therefore no resolved fit either). We cut out about 120 targets based on this criterion.

\subsection{Validation Against Published Far-IR Measurements}
\begin{figure*}
    \centering
    \includegraphics[width=\textwidth]{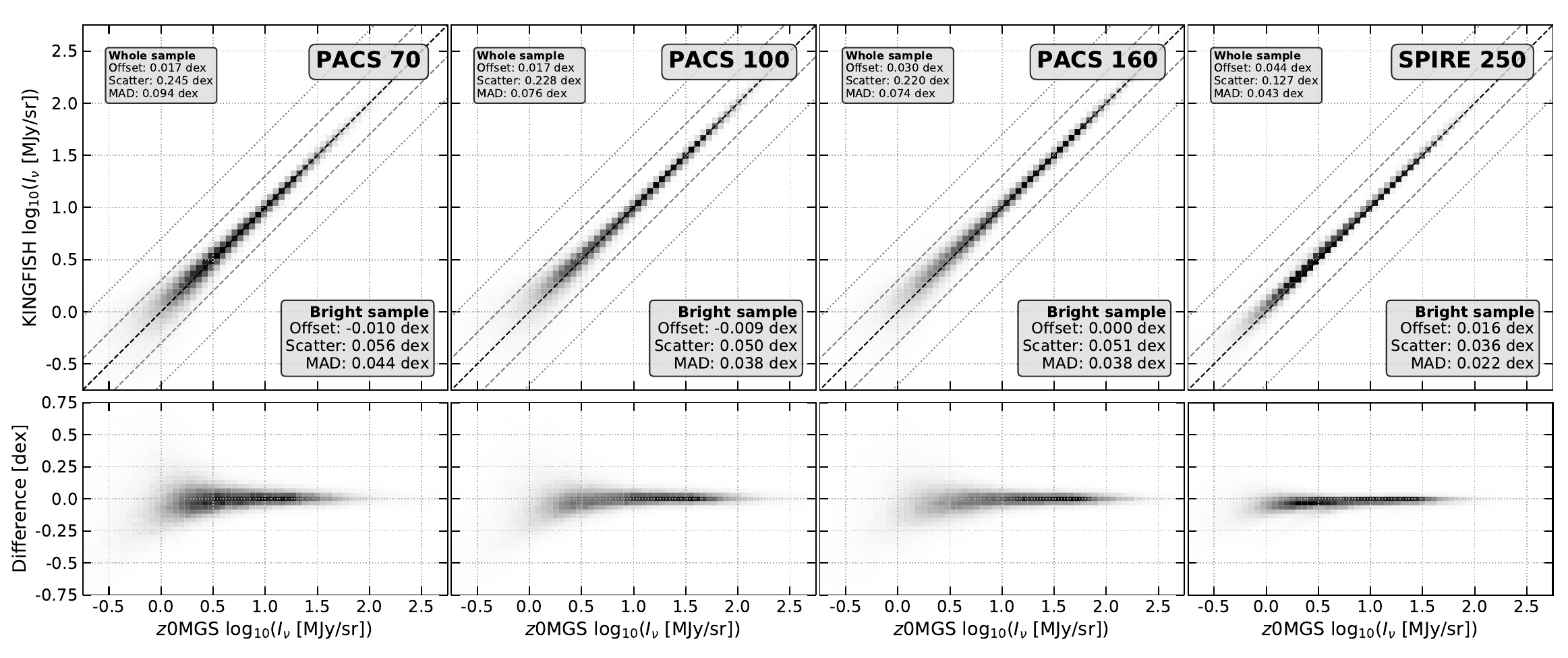}
    \caption{Resolved comparisons of surface brightnesses, in MJy~sr$^{-1}$, for overlapping galaxies between the KINGFISH \citep{Aniano20} and \textit{Herschel}-\zzmgs samples. 
    The top row shows the KINGFISH values as a function of the \textit{Herschel}\zzmgs surface brightness. The 1:1 relation is shown with black-dash line, and we add offsets lines (showing a factor of 2 and 5 difference) to guide the eye.
    The bottom row shows the difference (${\rm \textit{Herschel-}z0MGS - KINGFISH}$) in dex.
    We compute offset, scatter, and median absolute deviation (MAD) for the whole sample, shown in the upper-left corner of each panel, and for the ``bright'' pixels, chosen as $\log(I_\nu^{z0{\rm MGS}}~{\rm [MJy~sr^{-1}}]) \geq 0.75$, shown in the bottom-right corner of each panel.}
    \label{FigCompKINGFISHFluxes}
\end{figure*}

\begin{figure*}
    \centering
    \includegraphics[width=\textwidth]{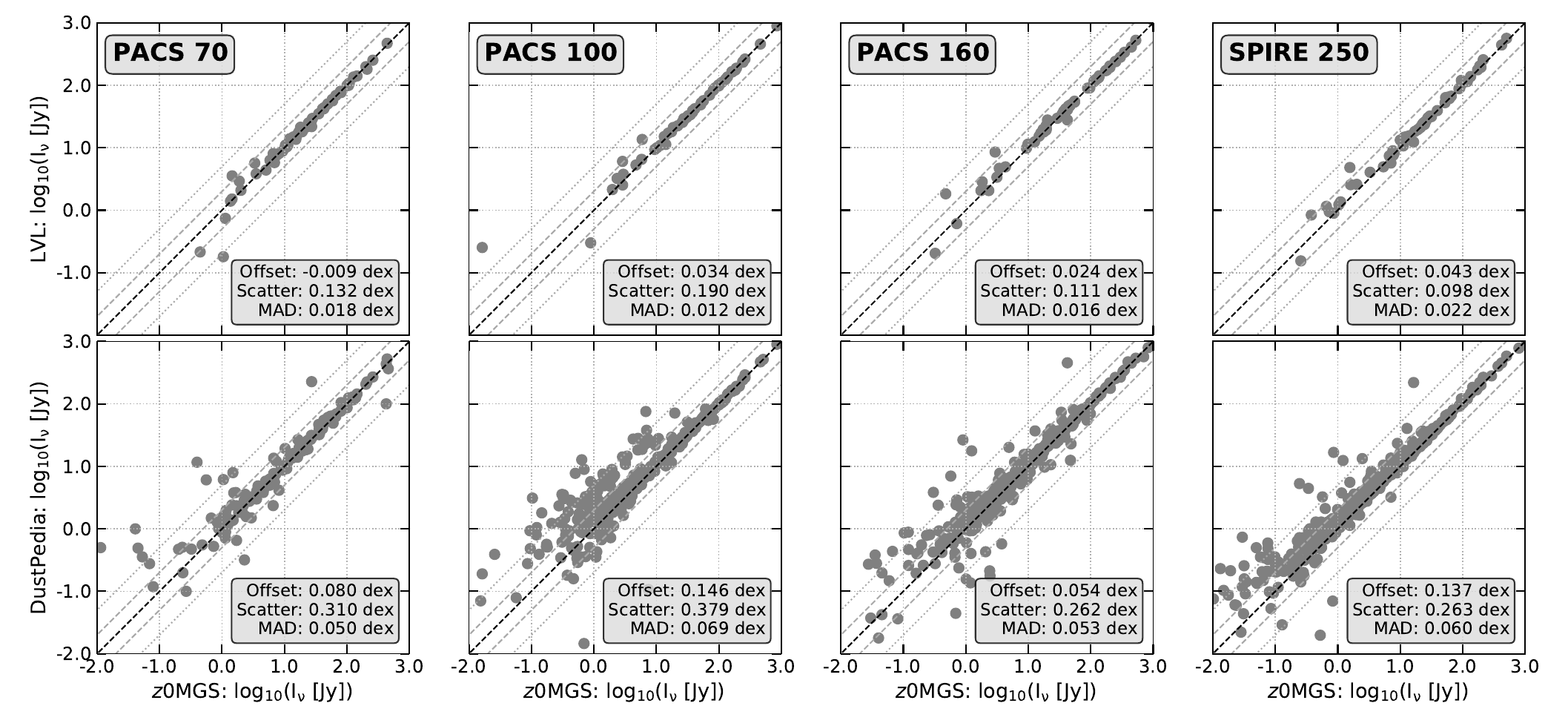}
    \caption{Comparisons of the integrated photometry from the \zzmgs sample compared to LVL (54 galaxies; top row), DustPedia (553 galaxies; bottom row). In general we find good agreement between the integrated photometry between our measurements and those of DustPedia and LVL, with the largest scatter for fainter galaxies. 
    Average noise values range between $2-4 \times 10^{-3}$~Jy in each band, corresponding to $\sim 0.51$, 0.36, 0.47, and 0.31~MJy~sr$^{-1}$ from PACS~70 to SPIRE~250.}
    \label{FigCompDPLVL}
\end{figure*}

In the following, we compare our PACS and SPIRE maps to previous measurements. Differences may arise from distinct pipeline processing, background subtraction (particularly at low surface brightness), and/or extraction apertures.  For each comparison we quantify the median offset between the fluxes or surface brightnesses, the 1$\sigma$ scatter, and the median absolute deviation (MAD), all in dex.  Figures~\ref{FigCompKINGFISHFluxes} and~\ref{FigCompDPLVL} present these comparisons.

\paragraph{KINGFISH}
We compare resolved surface brightnesses in our images to those measured by \citet{Aniano20}\footnote{The \citet{Aniano20} maps were obtained from \url{ http://arks.princeton.edu/ark:/88435/dsp01hx11xj13h}}. They investigated the dust properties in the KINGFISH galaxies \citep[][]{Kennicutt11KINGFISH}, using \textit{Spitzer} and \textit{Herschel} observations, working at the same SPIRE~250 resolution that we use for our SED fitting.  Nearly all KINGFISH galaxies are included in our sample, so the comparison of the resulting maps provides a one-to-one check on our procedures (NGC~5398 and NGC~5408 are not in the GALEX-WISE z0MGS sample, and we omit them here). In Figure~\ref{FigCompKINGFISHFluxes}, we reproject our maps onto the same grid as the \citet{Aniano20} images and compare the intensities between the two images pixel by pixel. The maps show outstanding agreement at high intensity, with little scatter, indicating that the processing yields consistent results with KINGFISH. The larger scatter at low intensities ($\log_{10}(I_\nu ~[{\rm MJy~sr^{-1}}]) \lesssim 0.75$) is due to the sensitivity of the KINGFISH maps, which is typically $5-7 \ {\rm MJy~sr^{-1}}$ for PACS and $\sim1 \ {\rm MJy~sr^{-1}}$ for SPIRE 250 \micron\ \citep{Kennicutt11KINGFISH}.

\paragraph{LVL} \citet{Dale2023} compiled spectral energy distributions, including \textit{Herschel} photometry for 
targets of the Local Volume Legacy survey \citep[][]{Dale2009LVL}. They conduct careful by-hand definition of local backgrounds and masking of contaminants like stars.
In the bottom row of Figure~\ref{FigCompDPLVL}, we compare their integrated fluxes and ours for 54 galaxies that overlap between samples. 
The overall agreement is excellent, with the integrated fluxes agreeing within a few percent on average and showing typically $0.01-0.02$~dex median absolute deviation. This overall good agreement with previous measurements confirms the validity of our data processing.

\paragraph{DustPedia}
We also compare the integrated fluxes measured within \rtf\ for our sample to integrated photometry for the DustPedia sample \citep[][]{Clark2018}, available from the DustPedia archive. 
An extensive description of the data processing is given in \citet{Clark2018}, with details on the CAAPR pipeline \citep[][]{Clark15, DeVis17a, DeVis17b}.
In Figure~\ref{FigCompDPLVL}, bottom row, we compare integrated fluxes between our work and DustPedia for 553 galaxies. The agreement between DustPedia and our sample is relatively good, though not as good as our agreement with the LVL and KINGFISH work.
We have visually checked the background for galaxies showing a deviation greater than 50\% between the two samples in PACS~100, and found no noticeable artifacts. The scatter is likely coming from the different data reduction approaches and differences in the integrated photometry aperture radius.
Although the MAD statistics are not as low as with the LVL sample, they are still within a reasonable range, and we consider that the agreement with these two samples is overall satisfactory.

\subsection{Data from the \zzmgs sample}
\label{SecZzmgsData}

\subsubsection{Integrated values}
We compare our observed dust and radiation field properties to galaxy-integrated values of the star formation rate (SFR), stellar mass (\stm) and offset from the star forming main-sequence (\dms) adopted from the \zzmgs catalog \citep[][]{Leroy19, irsa6}. The data were extracted from the \zzmgs delivery\footnote{\url{https://irsa.ipac.caltech.edu/data/WISE/z0MGS/overview.html}}. The definition of the star-forming main-sequence used to calculate the offset is given by Equation 19 in \citet{Leroy19}, reproduced here for clarity.
\begin{equation}
    \log_{10} ({\rm sSFR}_{\rm MS}) \ [{\rm yr}^{-1}] = -0.32 \log_{10} \left( \frac{M_*}{10^{10} M_{\odot}} \right) - 10.17
\end{equation}
For each integrated galaxy measurement we calculate the main sequence offset as:
\begin{equation}
    {\rm \Delta MS = log_{10}(sSFR) - log_{10}(sSFR_{MS})}.
\end{equation}

\citet{Leroy19} found that $\log($sSFR)$\leq -11$ should not be considered a robust estimate using the combination of UV and IR emission. In the following analysis, we only use targets above this minimum threshold.

\subsubsection{Galaxy-Averaged Surface Densities}
\label{sec:data_galaverage}
To bridge the integrated and resolved measurements, we also calculate ``galaxy-averaged'' surface densities by dividing the \stm and SFR by the galaxy's effective area. This calculation uses measurements of effective radii ($r_e$) from Sun et~al. (in prep) for the \zzmgs sample, including inclination corrections. We divide the integrated SFR and \stm from the \citet{Leroy19} atlas by the effective area ($\pi r_e^2$) for each galaxy where an $r_e$ value is available (176 galaxies do not have $r_e$ measurements).
We refer to these values as \gasfr and \gamstar.

\subsubsection{Surface densities}
\label{sec:surfacedensities}
We compute resolved maps of the star formation rate surface density (\sigsfr ), stellar mass surface density (\sigstm ), and resolved specific star formation rate, sSFR$_{R}$, using the prescriptions in \citet[][]{Leroy19}. To compute the stellar mass surface density, \sigstm, we use the estimated galaxy-integrated WISE~1 mass-to-light ratio $\Upsilon^{3.4}_*$ for each target provided in the \zzmgs catalog and 
\begin{equation}
    \Sigma_* \approx 3.3 \times 10^2 ~ \left ( \frac{\Upsilon_*^{3.4}}{0.5} \right ) ~ I_{3.4~\mu m}
\end{equation}
where \sigstm, is in \msolpcsq, and the surface brightness of the WISE~1 map, $I_{3.4~\mu m}$, is in MJy~sr$^{-1}$. While the true appropriate mass-to-light conversion may vary from location to location within a galaxy, the choice to use a single galaxy-integrated value ensures that the integrated galaxy results and the resolved results yield consistent stellar masses.

To compute \sigsfr, we use the WISE~4 photometry combined with GALEX FUV or NUV (with the top priority being FUV, next NUV if FUV is not available, and WISE~4 alone otherwise), per the hybrid prescription in \citet[][]{Leroy19}:
\begin{equation}
\begin{split}
    \Sigma_{\rm SFR} \approx f_{\rm FUV, NUV} ~ \left ( \frac{C_{\rm FUV, NUV}}{10^{-43.35}} \right ) ~ I_{\rm FUV, NUV} + \\
    3.24\times10^3 ~ \left ( \frac{C_{\rm WISE~4}}{10^{-42.7}} \right ) ~ I_{\rm 22~\mu m},
\end{split}    
\end{equation}
with $I_{\rm FUV, NUV}$, $I_{\rm 22~\mu m}$ in MJy~sr$^{-1}$,  $f_{\rm FUV}=1.04\times 10^{-1}$, and $f_{\rm NUV}=1.05\times 10^{-1}$. We refer to Table~7 in \citet[][]{Leroy19} to find the $C$ coefficients to use to convert the GALEX FUV or NUV and WISE~4 maps to \sigsfr in \msolkpcsqyr. 

We calculate the resolved specific star formation rate as \ssfrr $=$ \sigsfr/\sigstm.  To compute the resolved maps of the offset from the main sequence (\dmsr), we first define the resolved star forming main sequence using the maps of \sigstm and \ssfrr. We fit a line to binned averages of all pixels above the completeness thresholds (see Section~\ref{SecPixelsUsed}). 
We find ${\rm log_{10}({sSFR_{MS,R}}) = -0.625\times log_{10}(\Sigma_\star) - 5.216}$. 
We use that resolved main sequence to then calculate the offset from that prediction of all pixels based on their resolved specific star formation rate, as:
\begin{equation}\label{eq:resolved_ms}
    {\rm \Delta MS_R} = \log_{10}({\rm sSFR_R}) - \log_{10}({\rm sSFR_{MS,R}}).
\end{equation} 
We note that there are a variety of methodological choices possible for defining a resolved main sequence, so Equation~\ref{eq:resolved_ms} should be viewed primarily as a characterization of the offset from the average \ssfrr at a given \sigstm specifically for our sample.

\section{Dust emission modeling}
\label{SecFitting}
We fit the WISE through \herschel 250 \micron\ SEDs of dust emission using the \citet{DL07} physical dust models for both resolved and integrated measurements of each galaxy in our sample. We work at SPIRE 250 \micron\ resolution because it provides sufficient coverage of the long wavelength dust SED, while also preserving angular resolution \citep{Aniano12, Chastenet21}. 
In Appendix~\ref{sec:appendix500vs250}, we show the dust parameter distributions from fits using IR data up to SPIRE~500 (x-axes) against those from fits up to SPIRE~250 (y-axes), from the work of \citet{Chastenet21}. This shows that trends and parameter values are well kept without using the last two SPIRE bands.

\subsection{The \citet{DL07} model}
We use the \citet{DL07} model with the dust mass renormalization factor of 3.1 determined by \citet{Chastenet21}, which results from fitting the Milky Way high latitude cirrus spectrum from \citet{Compiegne2011} with depletion constraints from \citet{Jenkins2009} and \citet{Gordon14}. We use the \citet{DL07} MW (``Milky Way'') model, with $R_V=3.1$. Our procedure follows that of \citet{Chastenet21} exactly, and all details in this section can also be found there in more detail.

The radiation field heating the dust is described by a model with a delta function at a minimum radiation field, \umin, and a power-law distribution of radiation field intensities extending to $U_{\rm max}$ \citep{Dale01}.  The radiation field intensity is scaled in units of the \citet{Mathis83} Solar neighborhood radiation field intensity, $U_{\rm MMP}$.  The delta function plus power-law model has been used in a variety of dust SED modeling approaches, particularly with the \citet{DL07} model \citep[e.g.][]{Draine07,Aniano12,Aniano20}.
In this model, a fraction $(1-\gamma)$ of the dust grains is illuminated by \umin, while the fraction $\gamma$ is heated by the power-law distribution with $U_{\rm min} < U \leq {\rm U_{max}}$, defined as
\begin{equation}
\begin{split}
    \frac{1}{M_{{\rm d, }tot}} \left ( \frac{{\rm d}M_{\rm d}}{{\rm d}U} \right ) = 
    (1-\gamma)\ \delta (U - U_{\rm min})\ + \\
    \gamma \frac{\alpha - 1}{U_{\rm min}^{1-\alpha} -{\rm U_{max}^{1-\alpha}}}U^{-\alpha}.
\end{split}
\label{EquDustHeating}
\end{equation}
We fix the power-law coefficient to $\alpha = 2$ and we use ${\rm U_{max}} = 10^7$ \citep[following][]{Aniano12,Aniano20}. Using the fit results for \umin and $\gamma$, we also compute the average radiation field, \ubar:
\begin{equation}
    \overline{U} = (1-\gamma)\ U_{\rm min} + 
    \gamma \times U_{\rm min} \ \frac{{\rm ln(U_{max}}/U_{\rm min})}{1 - U_{\rm min}/{\rm U_{max}} }.
\label{EquUbar}
\end{equation}

We fit for the total dust surface density, \sigd, which combines both carbonaceous and silicate grain populations, and the mass fraction of carbonaceous grains with $<10^3$ carbon atoms, \qpah. 
We model the mid-IR starlight contribution to the SED (most notable at WISE 1 and 2) with a scaled 5,000~K blackbody, \omgs.

\subsection{Fitting procedure}
\label{SecFittingProcedure}
We use the grid-based Bayesian fitting code \dustbff \citep{Gordon14} to fit the dust emission model to the data. We use flat priors for all five parameters described in the previous section. Table~\ref{TabFitParams} lists the fitted parameters, with the sampling range and step size of the grid.

\renewcommand{\arraystretch}{1.1}
\begin{deluxetable}{llll}
\caption{Fitting parameters}
    \centering
    \tablehead{
    \colhead{Parameter} & \colhead{Range} & \colhead{Step} & \colhead{Unit}}
    \startdata
    \hline
    \hline
    \logt(\sigd) & [-3.0, 0.7] & 0.1 & \msolpcsq \\
    \qpah & [0, 6.1] & 0.25 & \% \\
    \umin & [0.1, 50] & Irr\tablenotemark{a} & -- \\
    \logt($\gamma$) & [-4.0, 0.0] & 0.15 & -- \\
    \logt(\omgs) & [-1.5, 3.0] & 0.1 & -- \\
    \enddata
\tablenotetext{a}{$U_{\rm min} \in $ \{0.1, 0.12, 0.15, 0.17, 0.2, 0.25, 0.3, 0.35, 0.4, 0.5, 0.6, 0.7, 0.8, 1.0, 1.2, 1.5, 1.7, 2.0, 2.5, 3.0, 3.5, 4.0, 5.0, 6.0, 7.0, 8.0, 10.0, 12.0, 15.0, 17.0, 20.0, 25.0, 30.0, 35.0, 40.0, 50.0\}.}
\label{TabFitParams}
\end{deluxetable}{}

\dustbff uses a multi-variate Gaussian with covariance matrix $\mathbb{C}$ to model the uncertainties due to statistical noise, background flux, and instrument calibration and their correlations. We use the same values as \citet[][]{Chastenet21} for the instrumental uncertainties. The background covariance matrix measures the correlation of the background signal for each band \citep[Eq. 23 in][]{Gordon14}. It changes for each set of mid- through far-IR images, and therefore we build a unique matrix for each galaxy (Section~\ref{SecMasks}). For fits to the integrated SEDs, the integrated fluxes would not be appropriately modeled with a background covariance matrix calculated with the same approach. In addition, the instrumental calibration uncertainties dominate over noise for the majority of galaxies. We therefore omit the background covariance matrix in this case. 

The fitting is done at the SPIRE~250 resolution ($\sim 18''$). We fit all pixels that are above 1$\sigma$ of the background in each band (not including WISE~1 and WISE~2), for a greater spatial coverage, and possible future use. However, we only use pixels above 3$\sigma$ of the background in each band (not including WISE~1 and WISE~2) for the following figures and analysis, as they offer more reliable results. 

\subsection{Integrated and Resolved fits}
We perform dust emission fitting on both integrated and resolved measurements.
Our final samples includes a total of \ngres targets with resolved fits. We also fit the integrated SED for an additional 58 targets that were not suited for pixel-by-pixel fitting due to their low surface brightness but were still significantly detected. This brings the number of galaxies with integrated fits to \ngint.

\subsection{Pixels used in the analysis}
\label{SecPixelsUsed}
We performed fitting on all pixels that passed a $1\sigma$ cut in all of the relevant infrared bands ($\sim 734,800$ total pixels corresponding to $\approx 41,000$ independent resolution elements).  For further analysis we make a series of stricter cuts to select high confidence dust emission measurements.  We also make several quality cuts to remove data with artifacts and poor fits from the sample. These cuts include:
\begin{itemize}
    \item \textit{S/N$>3$ in infrared Bands ---}
    Although the fit is performed on all pixels detected at $1\sigma$ above the background in all bands from WISE~3 to SPIRE~250, we only use pixels passing a $3\sigma$ cut in all bands in the following analysis. This ensures reasonable confidence in the SED fitting results for each pixel. Out of the original sample, the number of pixels that pass this cut is $\sim 474,900$ corresponding to $\approx 26,000$ independent resolution elements.
    \item \textit{Saturation in WISE ---}
    For bright nuclear point sources and for some very high surface brightness starburst galaxies, the WISE observations can be saturated. The \zzmgs catalog provides information on saturation in the WISE bands for their whole sample. We use their surface brightness thresholds (at the $7.5''$ resolution), 100~MJy~sr$^{-1}$ in WISE~1 and WISE~2 and 300~MJy~sr$^{-1}$ in WISE~3 and WISE~4, to cut out pixels exceeding these values, which are likely to experience saturation. We applied these ``saturation masks'' to NGC~598, NGC~1097, NGC~1365, NGC~3621, NGC~5236, and NGC~6946.
    Upon visual inspection, all other galaxies flagged with saturation in the \zzmgs catalog were discarded from the analysis due to having more extensive artifacts related to saturation. For IC~342, we also perform by hand masking, discarding pixels that show unrealistically high ${\rm sSFR_{R} > 10^{-8.5}~yr^{-1}}$, likely due to foreground star contamination. These cuts remove a total of $\sim 30\,000$ pixels ($2,000$ independent data points). 
    \item \textit{SED Fit Residuals ---}
    We only keep pixels that show fit residuals (described as {\tt (data - model) / model}) below 50\% in all PACS bands. This choice is discussed in further detail below in Section~\ref{sec:gammaissue}. This cuts out 240 pixels.
\end{itemize}

Unless mentioned otherwise, all subsequent figures only show pixels that pass these quality cuts.

\subsection{Radiation field parameters and limits of the fitting}\label{sec:gammaissue}
The \citet{DL07} models span a range of possible radiation field properties, with \umin ranging between 0.1 to 50 times the U$_{\rm MMP}$ radiation field and a potential range of $\log(\gamma)$ from $-$4 to 0. While this range encompasses the vast majority of points we expect to encounter in nearby star-forming galaxies, there are some environments where the radiation field may not be well-represented by our models. 

In intense starbursts, where the radiation field on $\sim$~kpc scales can be $>50$ U$_{\rm MMP}$, the models cannot reproduce the pervasive, high, minimum radiation field intensity. In such cases, we find that the fit results suggest a best fit where more of the dust luminosity is shifted into the power-law distribution. In this case, we typically see an underprediction of the 70 to 100~$\mu$m emission, due to the inability of the models to extend to \ubar values that generate SEDs peaking at $70-100~\mu$m. Such issues manifest as negative residuals from the fit in the PACS 70 and 100 bands, and inspection shows they are correlated with high fit $\gamma$ values.  

At the other end of the radiation field intensity range, it is important to note that we only include observations out to 250 \micron\ in our fit.  This means that for low radiation field intensities, where the peak of the thermal dust emission moves close to or beyond 250 \micron, we have a poor handle on the dust temperature and therefore dust mass, leading to bad fits.  Selecting galaxies with $\langle \overline{U} \rangle \sim \langle U_{\rm min} \rangle \leq 0.3$ (corresponding to $T \sim 15~$K), at the low end of our range of available \umin, all show large residuals in PACS~100, PACS~160, and SPIRE~250.
This suggests that these galaxies have dust cold enough that the fit up to 250~$\mu$m cannot reproduce the SEDs with the available \umin in our models (lower limit of \umin is 0.1). Additionally, we note that some of these galaxies do not have PACS~70 data. 

We find that both of these failure cases can be flagged via a cut on residuals in the PACS bands. 
We find that a cut on the residuals in the PACS bands at {abs(\tt (data - model) / model}) $>0.5$ is effective at isolating fits that show either very high or very low radiation field intensities with poor fits.  This cut has a minimal impact on the overall sample, flagging $240$ pixels overall, and we have verified that the main effect is to eliminate outliers and none of the scaling relationships change significantly if we tighten this cut to 25\% or relax the cut entirely.

\subsection{Completeness}
One of the goals of the analysis we present below is determining scaling relationships between \sigstm, \sigsfr, \sigd, \qpah, and the radiation field properties in our galaxies. In order to accurately measure these relationships, we first need to ensure that our sample is not biased due to the detection limits for {\em Herschel} and WISE, for the resolved scaling relations.
Because dust emission generally gets fainter as a function of galactocentric radius, our 3$\sigma$ cut translates into a cut in \sigstm and \sigsfr as well. This means that at lower values of \sigstm and \sigsfr, we will be biased towards selecting positive outliers in dust emission, either regions with more dust for a given combination of \sigstm or \sigsfr, or regions where noise pushes points above ${\rm S/N=3}$. This results in an overall bias in what we judge to be the average properties of the dust and radiation field. 
To quantify and minimize that bias, we calculate completeness thresholds in \sigsfr, \sigstm, \ssfrr, and \dmsr. To do this, we use the distribution of \textit{all} the pixels in the resolved maps prior to any cuts, and bin them by \sigstm, \sigsfr, \ssfrr and \dmsr.
In each bin, we calculate the fractions of pixels that pass the 3$\sigma$ cut in bands WISE~3--SPIRE~250 (note that the other cuts for saturation and residuals affect only a small fraction of the pixels so do not matter in judging completeness). We then find the value of \sigstm, \sigsfr, \ssfrr, and \dmsr where 50\% of pixels in a bin are below this 3$\sigma$ cut in any of those bands, we define this bin as our completeness threshold. 

From the results of this analysis, we consider coverage of the sample complete if:
\begin{itemize}
\setlength\itemsep{-0.2em}
    \item \logt(\sigsfr [M$_\odot$~yr$^{-1}$~kpc$^{-2}$])~$>-2.57$,
    \item \logt(\sigstm [M$_\odot$~pc$^{-2}$])~$>1.82$,
    \item \logt(\ssfrr [yr$^{-1}$])~$>-10.39$,
    \item \dmsr [dex] $>$ $-$0.29.
\end{itemize}
These thresholds are marked with vertical white-dashed lines in resolved-fits figures (showing quantities as a function of \sigstm, \sigsfr, \ssfrr, or \dmsr).
The analysis and subsequent scaling relations are inferred using only the pixels above these completeness thresholds. 
This corresponds to $\sim$~284,000, 180,000, 285,000, and 295,000 pixels above the thresholds in \sigsfr, \sigstm, \ssfrr, and \dmsr respectively.

\subsection{Resolved fits: comparison with KINGFISH}
\begin{figure*}
    \centering
    \includegraphics[width=\textwidth]{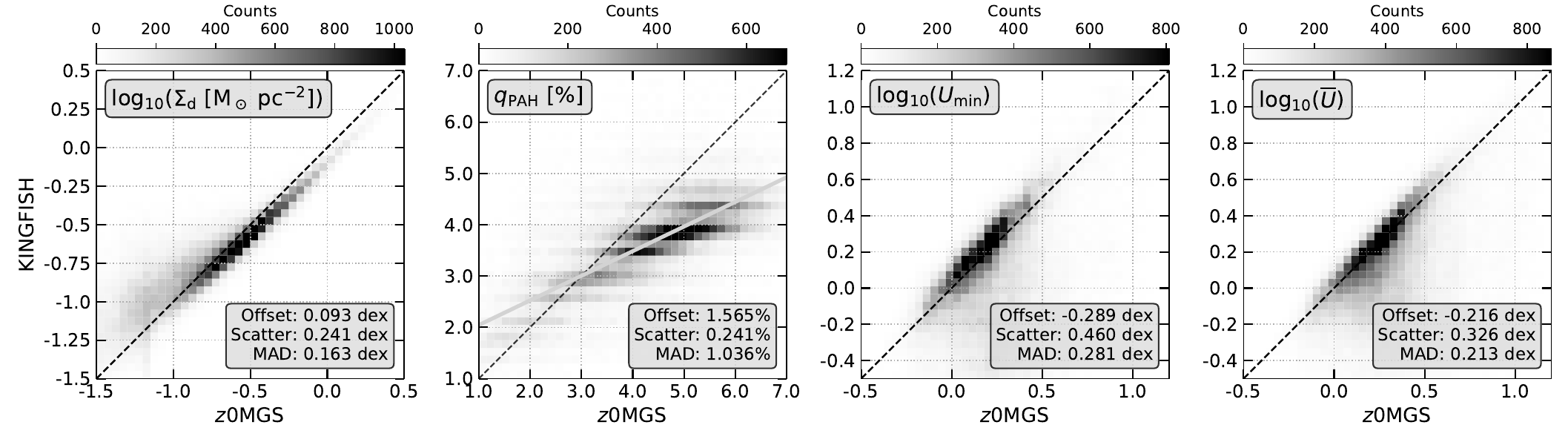}
    \caption{Comparison of dust and radiation field parameters between \textit{Herschel}-\zzmgs and \citet{Aniano20} for the KINGFISH galaxies (without NGC~5398 and NGC~5408). There is a good agreement between the two samples, despite different fitting approaches. The parameters and units are shown in each panel, with the offset, scatter, and median absolute deviation measured with respect to the 1:1 line (black dashed line), except for \qpah (second panel), where we find a best-fit slope of 0.49, shown in grey.
    Note that the \sigd distribution crosses the 1:1 line, which leads to a different offset if we measure it below or above a threshold of $\log_{10}(\Sigma_{\rm d}^{z{\rm 0MGS}}) = -0.6$.} 
    \label{FigCompKINGFISHResults}
\end{figure*}

In Figure~\ref{FigCompKINGFISHResults}, we compare our resolved results to those from \citet[][]{Aniano20} obtained fitting the \citet{Draine07} models to the KINGFISH sample, also working at SPIRE 250 \micron\ resolution.

The dust surface densities (\sigd) agree well between the two datasets, with a slight non-linearity in the slope. This likely reflects the adjusted dust opacities that we adopt.  In this work we adopt dust mass re-normalization factors for the \citet{DL07} models that were derived in \citet{Chastenet21}. Compared to the original \citet{DL07} model, the re-normalization resulted in a factor of 3.1 lower dust mass with no dependence on \umin or other radiation field parameters.  In \citet{Aniano20}, a different set of corrections were applied to the \citet{DL07} model based on \sigd offsets measured by comparing all-sky WISE and Planck SED modeling to extinction measured towards background quasars \citep{planck_aniano}.  The correction factor derived in that work included a \umin\ dependence, and the median value over the KINGFISH sample results in 1.6 times lower dust mass.  Thus we might expect to find a factor of approximately $3.1/1.6 \sim$2 lower dust masses in our results based on these different opacity assumptions, with closer agreement at the \umin\ values less than 1, where the \citet{Aniano20} correction factor increases.  This generally agrees with what we observe in the comparison of the \sigd results in the left panel of Figure~\ref{FigCompKINGFISHResults}, where our results lie below the one-to-one line at the highest \sigd (which tend to occur in inner regions of galaxies where \umin is higher) and get closer to the one-to-one line as \sigd decreases.  Thus, we argue that our results are in good agreement with \citet{Aniano20} given the different opacity assumptions. 

In the two right panels of Figure~\ref{FigCompKINGFISHResults}, we show the comparison between \umin and \ubar between our results and KINGFISH.  Here we also see good agreement.  Both \ubar and \umin are offset to 0.1$-$0.2 dex lower values in our fits relative to KINGFISH.  The fit in \citet[][]{Aniano20} allows the slope of the radiation field power-law distribution, $\alpha$, to vary, while we fix it to $\alpha=2$. While they find that on average $\alpha$ is consistent with $\sim2$, they do find small systematic variations between the inner and outer disks of some galaxies, with $\alpha$ being lower in the centers.  The fact that our \ubar\ and \umin\ are in good agreement suggests that allowing $\alpha$ to vary does not have a significant impact on the fitting and allows us to more efficiently explore the grid space of other fit parameters.

The largest differences in comparison to KINGFISH occurs in \qpah. We find that our results yield higher values of \qpah by $\sim 1$\% at the highest \qpah.  At lower values of \qpah, our results are in better agreement with KINGFISH.  These differences likely arise from the fact that we use mid-IR constraints from WISE rather than \textit{Spitzer}.  In general, the WISE 1 and 2 bands (i.e., 3.6 and 4.5 \micron) are very comparable to the \textit{Spitzer}/IRAC 1 and 2 bands and the \textit{Spitzer}/MIPS 24 \micron\ is a good match for WISE 22 \micron\ (WISE 4).  
The primary difference in replacing \textit{Spitzer} with WISE is therefore the shift of the main constraint on PAHs from the 7.7 \micron\ complex (sampled by the \textit{Spitzer}/IRAC 8 \micron\ filter) to the 11-12 \micron\ complex (sampled by the WISE 12 \micron\ filter).  The WISE 12 \micron\ filter also samples a wider wavelength range than \textit{Spitzer}/IRAC 8 \micron\ and therefore includes more continuum.  The \citet{DL07} model includes a fixed PAH spectrum with no variation in the intrinsic band ratios.  It may be the case that the difference in \qpah when using IRAC versus WISE could arise from our targets having brighter 11.3~\micron\ features relative to 7.7~\micron\ compared to what is in the model. In addition, many SINGS/KINGFISH galaxies show evidence for IRAC scan-direction artifacts in the \qpah maps from \citet{Aniano20}, which may reflect some persistence or saturation effects over the bright centers of galaxies.  The WISE maps are better behaved in this regard. In addition, the all-sky coverage of WISE also allows for a improved estimation of the background.  The \textit{Spitzer}/IRAC images in particular have relatively narrow fields of view, which could in some cases lead to over subtraction of the diffuse emission in the mid-IR, and underestimation of \qpah.  In general, the good correlation of the \qpah results from the two studies, despite their offset, give us confidence that our fits are tracing \qpah. Future work that incorporates both IRAC and WISE into the fitting or makes use of the more finely sampled mid-IR filter set from JWST may be able to reveal how PAH properties influence \qpah.  

To summarize, aside from a few minor differences and an overall scaling of \qpah, we find good agreement between our fitting results and those of \citet{Aniano20}. We proceed in the following sections to discuss the results from our SED fitting. 

\section{Results}
\label{SecResults} 

\subsection{Radiation Field Scaling Relations: Integrated Fits}
\label{sec:rad_int}
\begin{figure*}
    \centering
    \includegraphics[width=\textwidth]{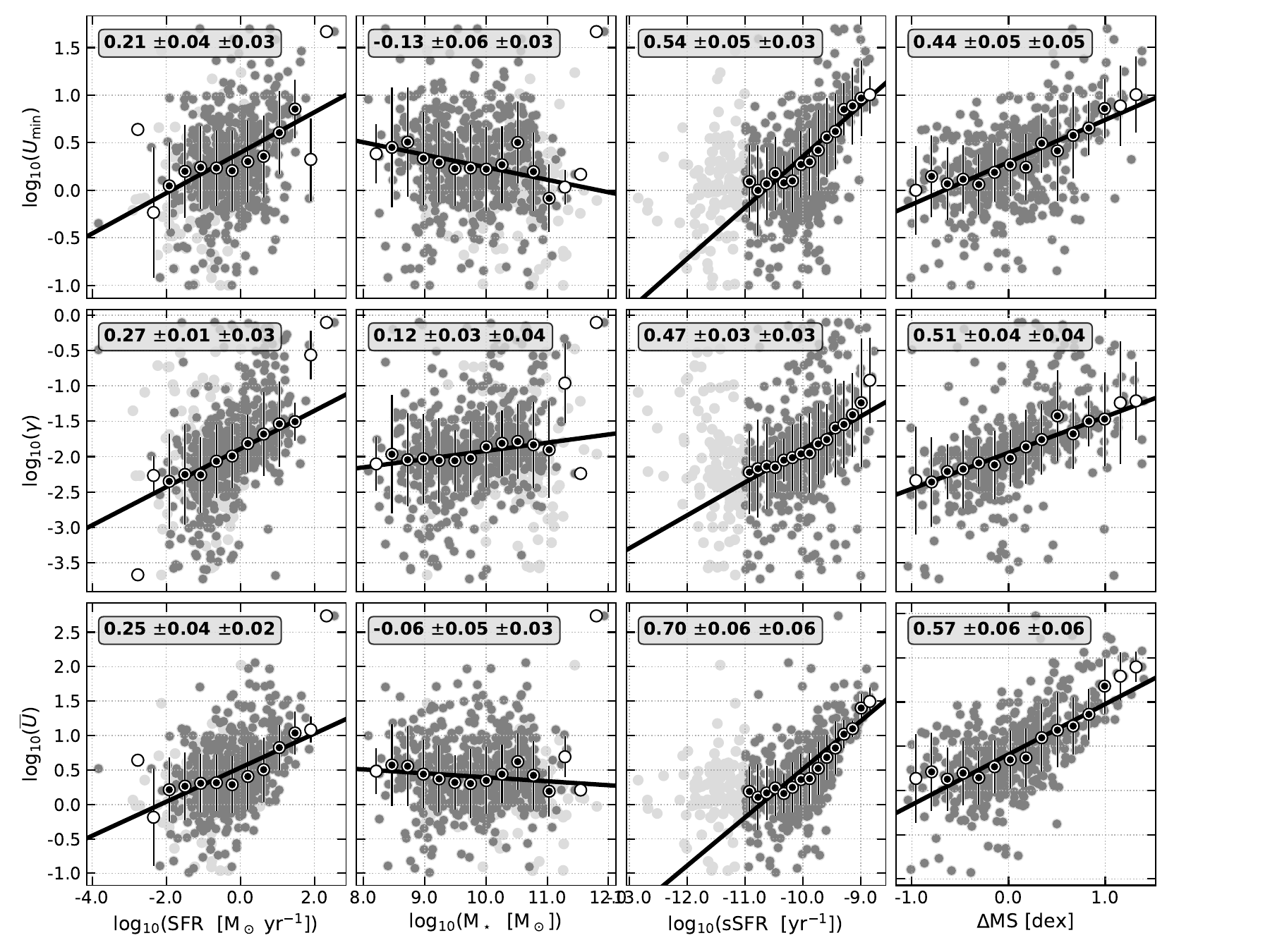}
    \caption{Distribution of \umin (top row), $\gamma$ (middle row), and \ubar (bottom row) from modeling integrated galaxies, as a function of sSFR (first column), SFR (second column), \stm (third column), and \dms (fourth column). Shown are all the integrated fits with acceptable residuals (which includes the light-grey points), and the darker points show those with \logt(sSFR)$>-11$. The white circles are the running medians (on dark points only), with error bars as one standard deviation. 
    The filled circles are those used to fit the scaling relations. The boxed text shows ${\rm a \pm \epsilon_{a}^{\rm fit} \pm \epsilon_{a}^{\rm bin}}$, with $Y = a\times X + b$, $\epsilon_{a}^{\rm fit}$ is the error on the slope for the shown fit, and $\epsilon_{a}^{\rm bin}$ is the standard deviation of slopes fit with a varying number of bins. These values are tabulated in Table~\ref{TabCoeffsFitsInt}.}
    \label{FigIntStarU}
\end{figure*}

\begin{figure*}
    \centering
    \includegraphics[width=\textwidth, clip, trim={0 7cm 0 0}]{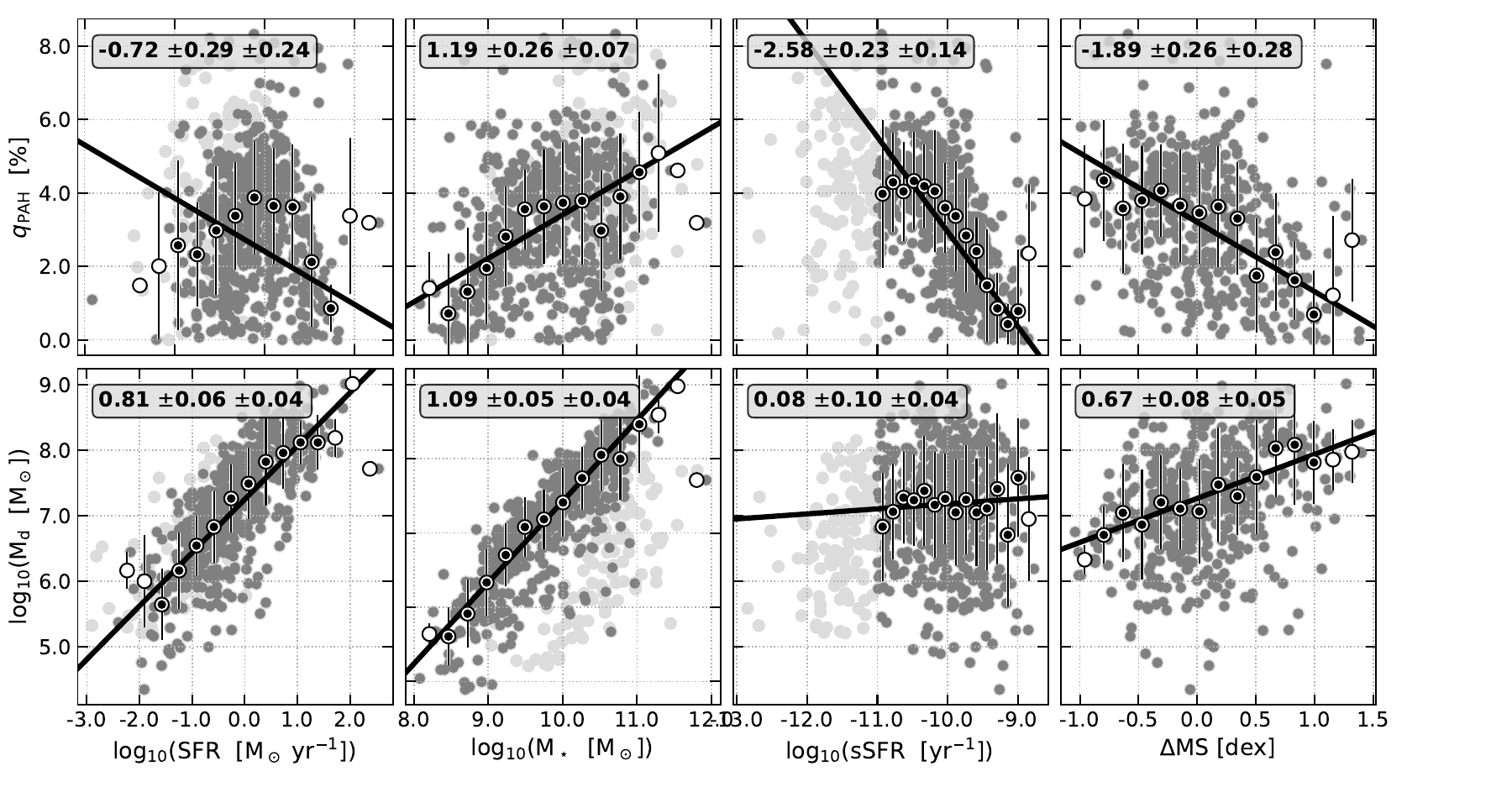}
    \caption{Same as Figure~\ref{FigIntStarU}, for fitted parameters \qpah (top row) and M$_{\rm d}$ (bottom row).}
    \label{FigIntStarDust}
\end{figure*}

We show the distribution of integrated dust and radiation field parameters \umin, $\gamma$, and \ubar as functions of SFR, ${\rm M_\star}$, sSFR, and \dms in Figure~\ref{FigIntStarU}, and similar scatter plots for \qpah and \mdust in Figure~\ref{FigIntStarDust}.
We show all fitted galaxies in light grey, and the darker points are galaxies with \logt(sSFR)$>-11$. 
In each panel, we show the binned medians considering only the dark points, with error bars showing one standard deviation. We estimate scaling relations by fitting the medians, where there are at least 10 points in a bin. We quote the results of power-law fits in Table~\ref{TabCoeffsFitsInt}\footnote{Throughout the paper we report power-law fit parameters for each scaling relation.  We note that there is no fundamental reason to expect power-law behavior between these variables, and in several cases there is evidence for variations that are not captured by our fits.  As an example, the \ubar-sSFR correlation in Figure~\ref{FigIntStarU} shows a relatively flat distribution at $\log({\rm sSFR}) \lesssim -10$ and a steeper slope above. We argue that to first order the power-law provides a straightforward and reasonably accurate representation of the data.}.  We note that in general, \umin and \ubar behave very similarly, indicating that the average interstellar radiation field is dominated by the \umin component, not the power-law distribution.

These plots reveal several trends:
\begin{itemize}
    \item All radiation field parameters (\umin, \ubar, $\gamma$) are positively correlated with quantities involving star formation, i.e., SFR, sSFR, and \dms. 
    \item The correlations with sSFR show the steepest slopes for all radiation field parameters. \umin and \ubar both change by around one order of magnitude or more as sSFR increases from $10^{-11}$ to $10^{-9}$ yr$^{-1}$.
    \item Because the ``main sequence'' is not flat in \stm--sSFR space, it is interesting to check for trends perpendicular to the main sequence at a given \stm, quantified by \dms.  In the right panels of Figure~\ref{FigIntStarU} we show that the trends of radiation field parameters \umin and \ubar with \dms show slightly shallower slopes than the correlation with sSFR.
    \item Correlations of \umin, \ubar, and $\gamma$ with SFR all show shallower slopes compared to trends with sSFR and \dms. 
    \item \umin, \ubar, and $\gamma$ have little correlation with \stm.  This is particularly notable since \stm spans $>3$ orders of magnitude over the completeness range of our sample. 
    \item Judged by the root mean square error (RMSE) of the correlation, \ubar, \umin, and $\gamma$ can be best predicted using \dms, though SFR also predicts $\gamma$ with nearly the same RMSE. Note that calculating \dms requires knowing \stm\ and SFR.
\end{itemize}

In general, these findings agree with expectations for how the ``delta-function plus power-law model'' should behave, and give insights into what is setting the \umin value, which is the dominant part of the ISRF.  In theory, both recent star formation and existing older stellar populations (B stars and later) should contribute to setting \umin. The power-law distribution is expected to arise from dust heated by higher intensity radiation fields in the vicinity of recently formed stars.  The correlation of $\gamma$ with SFR, sSFR, and \dms agrees with this picture.  Interestingly, \umin (and \ubar) correlations show little dependence on \stm, which may indicate that current star formation dominates the radiation field on integrated scales. However, the correlation of \ubar and \umin with SFR is also fairly weak, which could be interpreted as extensive quantities like \stm and SFR not capturing the physics that set the radiation field (e.g., the radiation field distribution should differ between a compact, high intensity distribution of SF compared to a larger, low intensity distribution).  These integrated relationships may also simply reflect a different spatial distribution of the stellar mass, star formation, and ISM throughout the galaxy, for example if stellar mass is highly centrally concentrated while dust and star formation are more distributed through the disk.  We address this topic in Sections~\ref{sec:rad_res} and~\ref{sec:galavscalings} by looking at the resolved and galaxy-averaged versions of these plots.

\input{tab_integratedfits}

\subsection{Radiation Field Scaling Relations: Resolved Fits}\label{sec:rad_res}
\begin{figure*}
    \centering
    \includegraphics[width=\textwidth]{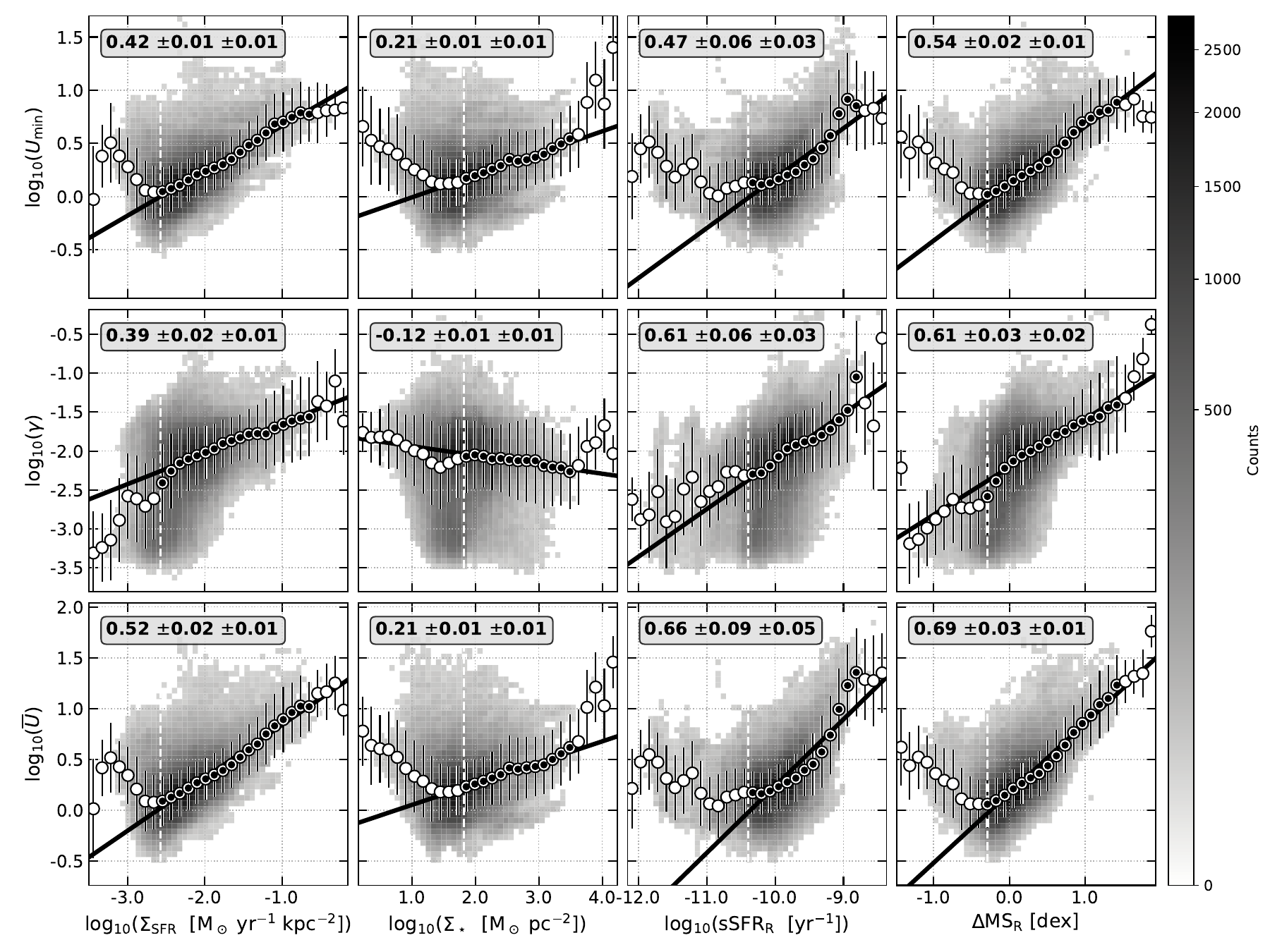}
    \caption{2D histograms of \umin (top row), $\gamma$ (middle row), and \ubar (bottom row) fit for individual lines of sight at the SPIRE 250$\mu$m resolution, as a function of \ssfrr (first column), \sigsfr (second column), \sigstm (third column), and \dmsr (fourth column), color-coded by the number of hits. 
    The vertical white-dashed lines are the limits of completeness (\S \ref{SecPixelsUsed}). 
    The white circles are the running medians, with error bars as one standard deviation, and the filled circles are those used to fit the scaling relations. We mark the slope in each panel.}
    \label{FigStarU}
\end{figure*}

\begin{figure*}
    \centering
    \includegraphics[width=\textwidth, clip, trim={0 7cm 0 0}]{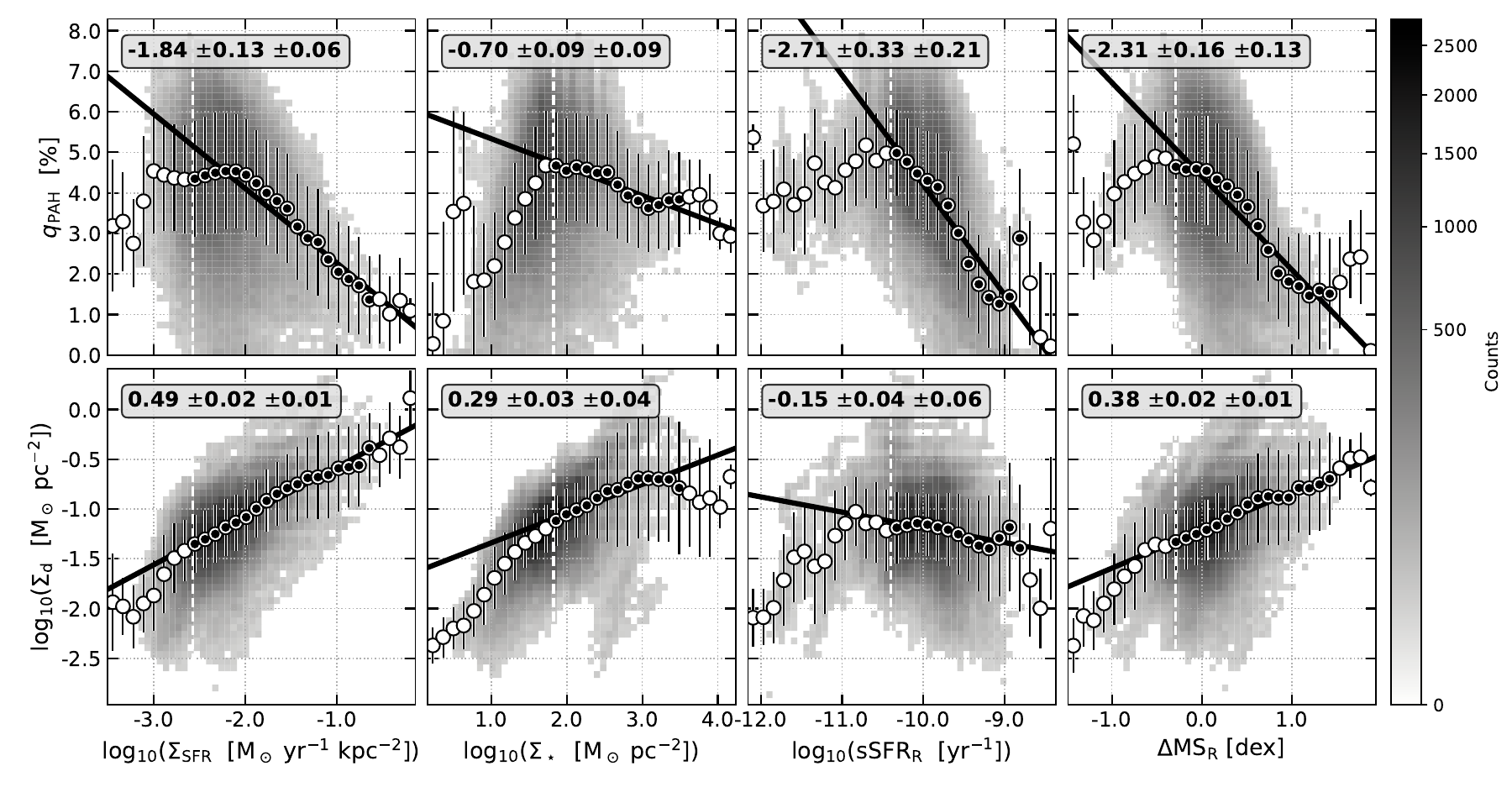}
    \caption{Same as Figure~\ref{FigStarU}, for fitted parameters \qpah (top row) and \sigd (bottom row).}
    \label{FigStarDust}
\end{figure*}

\begin{figure}
    \centering
    \includegraphics[width=0.5\textwidth]{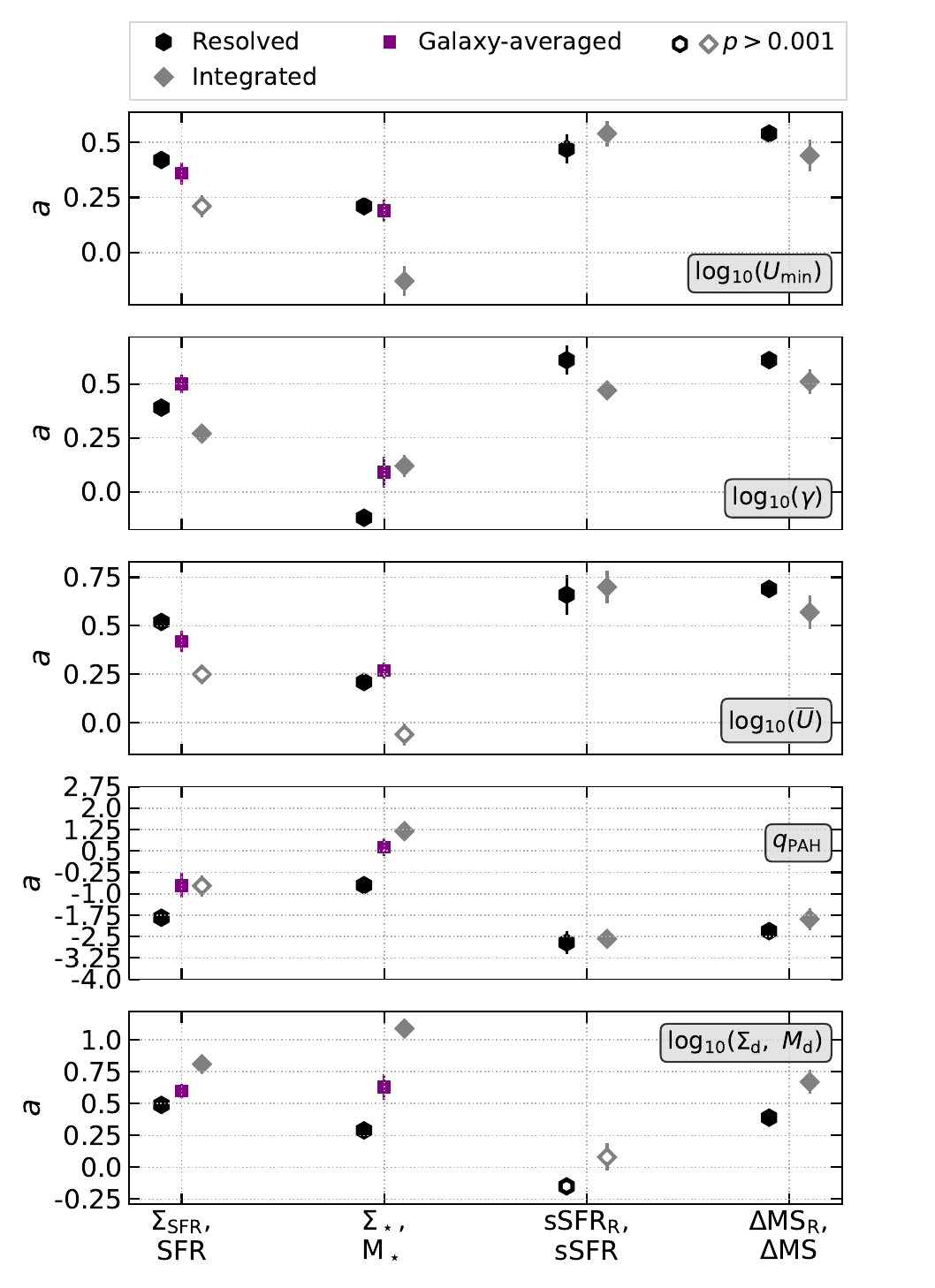}
    \caption{Comparison of the slopes ($a$) of resolved (black symbols) and integrated (grey symbols) fits presented in Figures~\ref{FigIntStarU}--\ref{FigStarDust} and Tables~\ref{TabCoeffsFitsInt} and~\ref{TabCoeffsFitsResolved}. We also show the fit to $\langle {\rm SFR} \rangle$ and $\langle M_\star \rangle$, galaxy-averaged values, shown in Figure~\ref{FigGalAveraged}.
    Here each fit follows the form $Y= aX + b$, where we show only $a$ in this Figure. The open symbols mark where the high $p$-value signals no significant correlation between the axes.  The behavior of the slopes of the \umin and \ubar relationships are very similar, since under most circumstances the radiation field is dominated by the minimum radiation field.}
    \label{fig:SlopesComp}
\end{figure}

In Figure~\ref{FigStarU}, we show the 2D density histograms of dust parameters \umin, $\gamma$, and \ubar as functions of \ssfrr, \sigsfr, \sigstm, and \dmsr, and similar histograms for \qpah and \sigd in Figure~\ref{FigStarDust}.
In each panel, we show the binned medians, with error bars showing one standard deviation. The filled circles mark the medians used to fit the power-law scaling relationships, selected to have at least 250 measurements in a bin.
We report the fits in Table~\ref{TabCoeffsFitsResolved} and also show a comparison between the slopes derived for the scaling relationships in the integrated and resolved fits in Figure~\ref{fig:SlopesComp}.
We note the excellent agreement between the completeness thresholds defined in Section~\ref{SecPixelsUsed}, computed independently, and the behavior of the binned medians, which begin to noticeably deviate from the trends below the completeness limits. 
In Appendix~\ref{app:corner} we present correlations between the resolved fit parameters (\sigd, \qpah, \umin, \ubar, and $\gamma$) to give further insight into their behavior.

\input{tab_resolvedfits}

The resolved scaling relations between radiation field parameters (\ubar, \umin, and $\gamma$) and \sigstm, \sigsfr, \ssfrr, and \dmsr reveal the following trends:
\begin{itemize}
    \item All radiation field parameters are positively correlated with quantities related to star formation: \sigsfr, \ssfrr, and \dmsr, similar to what was found for integrated measurements. 
    \item The steepest slopes occur in the relationships between the radiation field quantities and \ssfrr\ or \dmsr, suggesting specific star formation rate and related quantities are important for setting the radiation field distribution.
    \item Despite the lack of correlation with \stm\ in the integrated measurements, \ubar and \umin ~{\em do} show a significant correlation with \sigstm\ in the resolved measurements.  This could result from the older stellar distribution contributing to setting \umin , with \sigstm\ tracing the density of old stars, or it might reflect that \sigstm and \sigsfr are often both radially decreasing and therefore correlated.
    \item The fraction of dust luminosity from the power-law distribution, $\gamma$, does not strongly depend on \sigstm, a weakly negative correlation.  This reinforces that the older stellar mass distribution is not critical in setting the power-law part of the radiation field distribution.
    \item In general, radiation field parameters are slightly better predicted by \dmsr compared to \ssfrr. There is a small decrease in the RMSE for all relationships of radiation field quantities with \dmsr compared to \ssfrr, suggesting the perpendicular offset from the resolved main sequence is better at predicting the radiation field properties than \ssfrr alone. Given that our measurement of the resolved MS has a non-linear slope of -0.625 (see Eq~\ref{eq:resolved_ms}), \ssfrr and \dmsr are not simply scaled versions of each other, so the slight but consistent preference for \dmsr as a radiation field predictor suggest that this is meaningful.   
    \item Judged by the RMSE of the correlation, \ubar and \umin can be best predicted using \sigsfr or \dmsr.  $\gamma$ is best predicted with \dmsr. We note that some part of the RMSE can be due to behavior not captured by our power-law fits.
\end{itemize}

The difference in slope between the resolved and integrated fits potentially provides some insight into the drivers of the radiation field distribution in galaxies.  Figure~\ref{fig:SlopesComp} summarizes the slopes of the various scaling relationships as a function of integrated or resolved stellar mass, SFR, sSFR, and $\Delta$MS. Some trends we see in this comparison include:
\begin{itemize}
    \item The slopes of scaling relations for all radiation field quantities (\umin, \ubar, and $\gamma$) versus sSFR or \ssfrr remain consistent within their uncertainties between the integrated to resolved analysis.  This may be expected since sSFR is a ratio of surface densities, so may to first order remove radial variation of \sigstm and \sigsfr within galaxies that leads to differences in resolved and integrated fits. 
    \item The slope of $\gamma$ vs \sigstm and M$_*$ are slightly different, though both similar to a flat or slightly decreasing trend, indicating that on both integrated and resolved scales, $\gamma$ is not closely tied to the distribution of old stars.
    \item The behavior of \umin and \ubar are very similar, as would be expected. In both of these parameters, we see that the slope of the resolved relationships is steeper than the equivalent integrated scaling relationship.  
\end{itemize}

To summarize, the SED modeling results find a strong correlation between the minimum radiation field intensity, \umin, and the average radiation field intensity, \ubar, with quantities tied to SFR or sSFR.  \umin and \ubar also show correlation on resolved scales with stellar mass surface density, \sigstm, suggesting that both older stars and recent star formation contribute to setting the typical radiation field intensity.  The power-law component  does not show significant correlation with \sigstm, but does with \sigsfr, in agreement with the expectation that these more intense radiation fields are caused by proximity to star forming regions.  

The $\gamma$ parameter is weakly, positively correlated with \stm for integrated galaxies, which is unexpected given that it is tied to higher radiation field intensities. The correlation we observe may be due to the fact that the power-law plus delta function model applied to the entire galaxy enforces that a single \umin\ must represent all areas of the galaxy. If \umin is tracing \sigsfr then the galaxy actually has a range of resolved \umin values.  The integrated \umin may lie at the low end of the distribution of resolved \umin values and push all higher \umin values into the power-law distribution. 

In general, \umin, \ubar, and $\gamma$ recovered by the power-law plus delta function model fits, particularly on resolved scales, appear in good agreement with expectations for the behavior of these components of the radiation field.  This is discussed further in Section~\ref{sec:disc_rad}.

For predicting the resolved average radiation field intensity, \ubar, the best single quantity is \sigsfr (followed closely by \dmsr). For $\gamma$ it is \dmsr and \sigsfr.  We note \dmsr as defined in Equation~\ref{eq:resolved_ms} is not trivially applied outside our dataset, so \sigsfr is likely the most accessible predictor for resolved radiation field quantities.  In Section~\ref{sec:planefits} we explore if the prediction of \ubar is improved using a combination of \sigstm and \sigsfr.

\subsection{Dust Properties Scaling Relations: Integrated Fits}
\label{sec:dust_int}

The results of integrated fits of the dust mass and PAH fraction are shown in Figure~\ref{FigIntStarDust}. Trends evident in \qpah in these plots include: 
\begin{itemize}
    \item \qpah is not strongly correlated with SFR but is strongly, negatively correlated with sSFR and \dms.  This result suggests that at a fixed \stm, galaxies with higher SFR have lower PAH fractions.  This may be tied to the destruction of PAHs in ionized gas, as discussed in Section~\ref{SecPAHFraction}. 
    \item \qpah is positively correlated with \stm.  This scaling relation may result from a combination of a metallicity trend (lower \qpah at lower metallicity plus the mass-metallicity relationship) and a sSFR-related destruction of PAHs in \hii\ regions (higher sSFR at lower \stm).
\end{itemize}
As discussed in Section~\ref{SecPAHFraction} our results compiling a large sample of resolved nearby galaxies provide key context for previous studies of resolved PAH fraction in smaller samples.  Our scalings also agree with higher resolution PAH fraction mapping which clearly show the lower PAH fractions in \hii\ regions likely due to destruction of PAHs \citep[e.g.,][]{Gordon2008, Paradis2011, Paladini2012, Relano2016, Chastenet19,Chastenet2023, Egorov2023, Sutter2024}. In Section~\ref{SecPAHFraction} we discuss these results further and compare to past literature results on these correlations.

The bottom row of Figure~\ref{FigIntStarDust} shows the integrated scaling relations of dust mass (\mdust) as a function of SFR, \stm, sSFR, and \dms.  These panels show a few important trends:
\begin{itemize}
    \item \mdust is well correlated with both SFR and \stm. We find a remarkably linear slope in \mdust--\stm and small scatter. The correlation with SFR is slightly weaker and sub-linear.   
    \item We find that \mdust is uncorrelated with sSFR, but shows a slight positive correlation with \dms. This suggests that the lack of sSFR trend could result from competing effects of lower dust-to-gas ratios for lower \stm galaxies and higher M$_{\rm gas}$/\stm for galaxies with positive offsets above the main sequence.
\end{itemize}
Trends of \mdust as a function of various galaxy integrated quantities have been explored in detail in the literature. In Section~\ref{SecDustandStellar}, we discuss our results in comparison to past measurements and scaling relationships.  

\subsection{Dust Properties Scaling Relations: Resolved Fits}

In Figure~\ref{FigStarDust} we show the resolved correlations between \sigd and \qpah as a function of \sigsfr, \sigstm, \ssfrr, and \dmsr. We note the following results about the behavior of \qpah:
\begin{itemize}
    \item The resolved PAH fraction decreases steeply with \sigsfr, \ssfrr, and \dmsr.  The steepest decrease occurs as a function of \ssfrr, where over the span of slightly more than 1 dex in \ssfrr, the \qpah drops from values $\sim4-5\%$, representative of the diffuse ISM of the Milky Way \citep{DL07} to $\sim1\%$ at $\log$\ssfrr$\sim-9$. This trend is less steep in \dmsr and \sigsfr, though both reach \qpah$\sim1\%$ at their highest values.  These trends suggest that the intensity of star formation in a given region is critical to setting the PAH fraction.
    \item \qpah is weakly correlated with \sigstm, with a negative slope.  This trend may result from the overall radial correlation of \sigstm and \sigsfr.  The highest \sigstm values tend to fall in galaxy centers, where \sigsfr can be high leading to lower \qpah. The decrease at high \sigstm could also be related to the less efficient excitation of PAH vibrational modes where older stars contribute significantly to the radiation field \citep{Draine2014,Draine21,Whitcomb2024}.
\end{itemize}
We note that this is the first compilation of resolved measurements of \qpah for a large, diverse sample of galaxies, where we can control for completeness effects in \sigsfr and \sigstm. Past studies of small samples have tended to select bright, highly star-forming galaxies that sit above the typical integrated or resolved star-forming main sequence.  This makes interpreting trends as a function of other quantities like metallicity subject to biases due to incomplete sampling of sSFR. Our results clearly show a steep, negative dependence of \qpah on quantities related to SFR, which is likely related to their destruction in the ionized gas of \hii\ regions.  We discuss these insights in more detail in Section~\ref{SecPAHFraction}.

The bottom row of Figure~\ref{FigStarDust} shows the resolved behavior of dust mass surface density, \sigd.  Here we note the following trends:
\begin{itemize}
    \item \sigd is positively correlated with \sigsfr, \sigstm, and \dmsr, but weakly, negatively correlated with \ssfrr.
    \item The \sigd--\sigsfr relationship can be interpreted as a reflection of the resolved Kennicutt-Schmidt relationship \citep{Bigiel08, Leroy08}, a topic we discuss further in Section~\ref{sec:inverseks}.
    \item \sigd is positively correlated with resolved main sequence offset, \dmsr and slightly negatively correlated with \ssfrr.  This reflects the fact that the resolved main sequence has a slope as a function of \sigstm.  The trend of higher \sigd at a fixed \sigstm for regions with higher \sigsfr (i.e., the positive correlation with \dmsr) goes in the direction one would expect if those regions have higher gas surface densities.  This agrees with resolved scaling relationships between gas, star formation rate, and stellar mass surface densities.
\end{itemize}

Comparing integrated and resolved results (Figure~\ref{fig:SlopesComp}), the most dramatic differences are in the dependence of \qpah on \sigstm and \sigsfr. In the resolved measurements, \qpah decreases strongly with \sigsfr, while the integrated \qpah is only weakly negatively correlated with SFR. We interpret this as a result of PAH destruction being related to the local intensity of star formation, which the resolved analysis better isolates. In the integrated analysis, \stm--metallicity and \stm--SFR scalings may overwhelm this signal.

\subsection{Scaling relations: Galaxy-averaged Surface Densities Fits}\label{sec:galavscalings}
The results of the integrated analysis presented in Sections~\ref{sec:rad_int} and \ref{sec:dust_int} are not straightforwardly comparable to the resolved measurements since they mix together ``intensive'' and ``extensive'' quantities, resulting in correlations that are more difficult to interpret. For example, \umin or \ubar are intensive, average-like values, \qpah and $\gamma$ are fractions, while SFR and \stm are extensive, sum-like integrated values. This means that a quantity like SFR can increase by adding together many low SFR regions, while \ubar or \qpah can stay the same under those conditions. This effect is likely to play a role in the different scalings we show in the resolved and integrated cases.
To investigate this issue, we consider galaxy-averaged SFR, \gasfr, and \stm, \gamstar, as described in Section~\ref{sec:data_galaverage}, to create intensive \stm and SFR quantities by dividing by the effective area of the galaxy.

\begin{figure}
    \centering
    \includegraphics[width=0.48\textwidth]{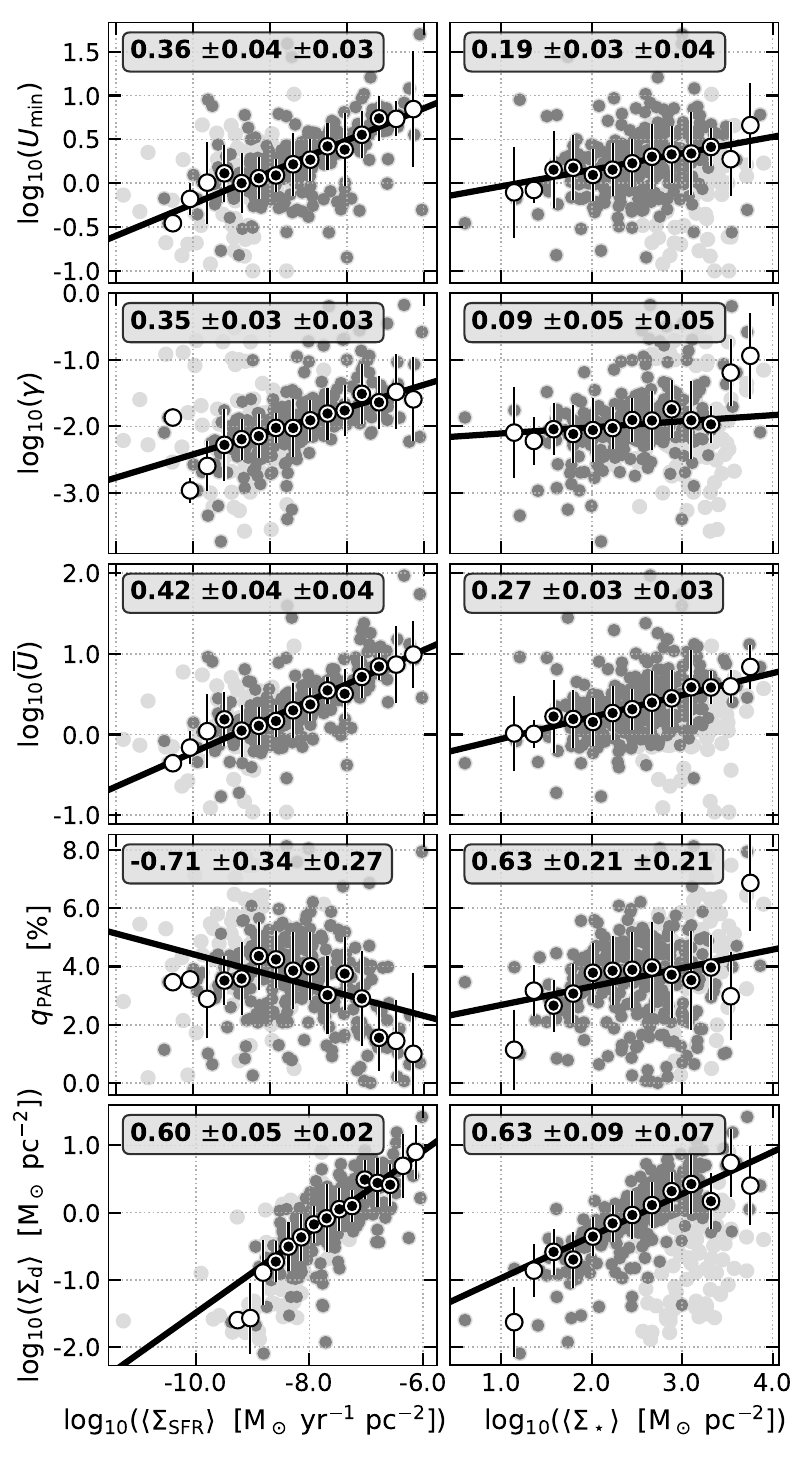}
    \caption{Galaxy-averaged SFR and \stm, $\langle {\rm SFR} \rangle$ (left) and $\langle M_\star \rangle$ (right), variations with the dust parameters \umin, $\gamma$, \ubar, \qpah, and galaxy-averaged dust mass, $\langle M_d\rangle$. The symbols show the binned medians and lines are linear fits to the filled medians, similarly to Figures~\ref{FigIntStarU}--\ref{FigStarDust}.}
    \label{FigGalAveraged}
\end{figure}

In Figure~\ref{FigGalAveraged}, we show the radiation field and dust parameters \umin, \ubar, $\gamma$ and \qpah as a function of the galaxy-averaged SFR and \stm, noted \gasfr and \gamstar. We also apply the same galaxy-averaged measurement to the total dust mass so that we look at the galaxy-averaged dust mass, \gamd. 
Here we see that most of the galaxy-averaged trends are very similar to the resolved trends, though the slopes are often slightly shallower. We report these slopes in Figure~\ref{fig:SlopesComp}.  Galaxy-averaged scalings may be of use particularly for distant galaxies, where estimates of $r_e$ are possible with HST and JWST imaging, but IR measurements may remain unresolved. Using the galaxy-averaged values provides a closer match to the resolved scaling behavior in most cases.

\subsection{Predicting Radiation Field and Dust Properties}\label{sec:planefits}
In the previous sections, we presented scaling relations between radiation field and dust parameters with single quantities related to stellar mass and star formation rate.  For the purposes of predicting these properties, it is interesting to determine whether combinations of these variables yield improved predictions.  

\begin{figure*}
    \centering
    \includegraphics[width=\textwidth]{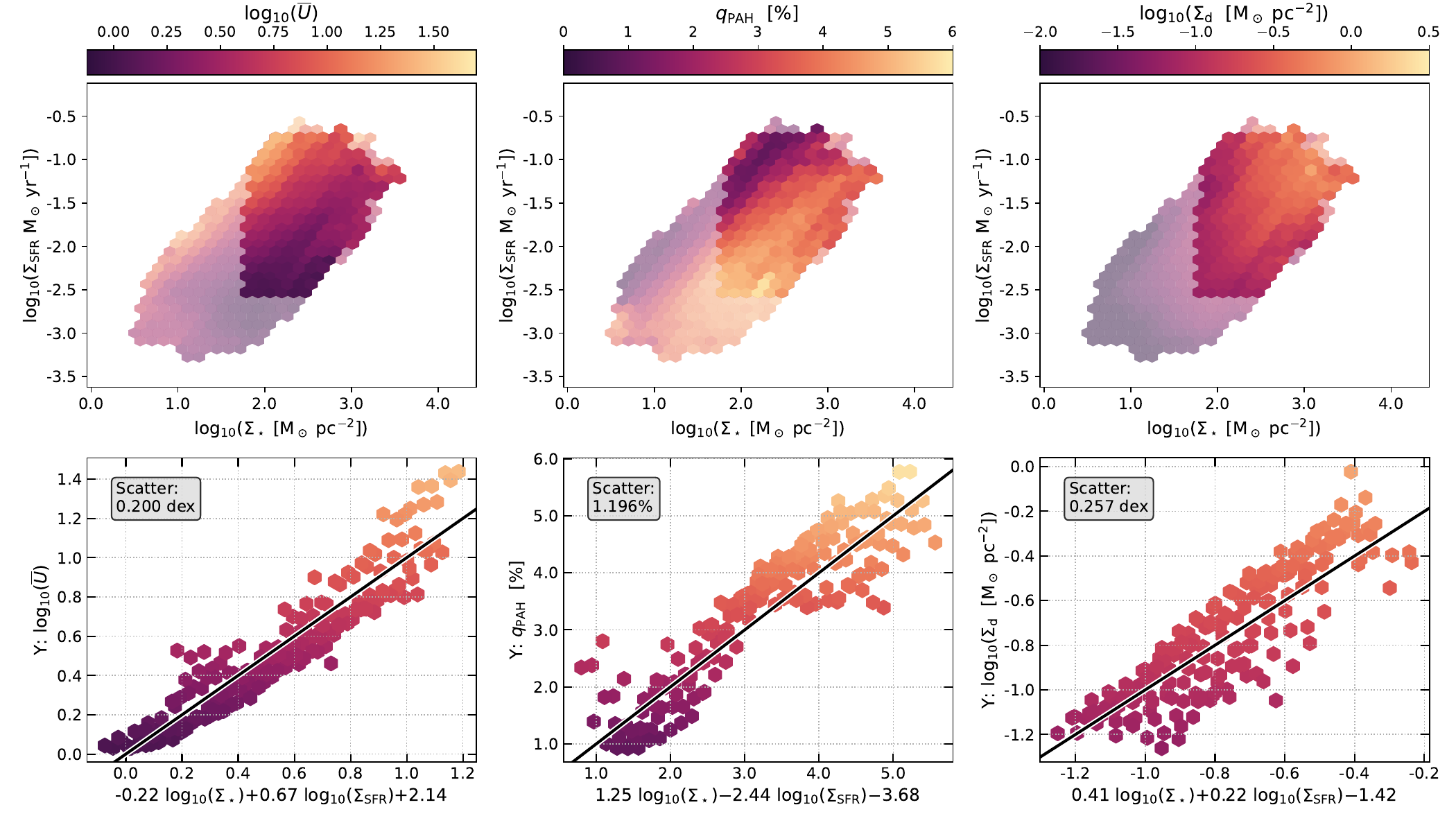}
    \caption{Two-dimensional fits of \ubar, \qpah, and \sigd as a function of a combination of the resolved stellar mass and star formation rate.
    \textit{Top row:} 2D histograms of the \sigstm--\sigsfr plane, where the bin values show the average \ubar (left column), \qpah (center column), and \sigd (right column). The light bins are calculated for all pixels passing the 3$\sigma$ S/N cut, and the bright bins are calculated for pixels passing that same cut \textit{and} that that have \sigstm and \sigsfr above the completeness thresholds. 
    The 2D fit uses the standard deviations within each bin as errors in the linear regression.
    \textit{Bottom row:} The x-axis shows the best-fit using both \sigstm and \sigsfr. The y-axis shows the observed data. The best-fit coefficients are reported in Table~\ref{TabCoeffs2d}.}
    \label{Fig2DFits}
\end{figure*}

In Figure~\ref{Fig2DFits}, we present two dimensional fits to a combination of \sigstm and \sigsfr to determine a prediction for \ubar, \qpah, and \sigd. The results of these fits are also summarized in Table~\ref{TabCoeffs2d}.
In the top row of Figure~\ref{Fig2DFits}, we show the two dimensional histograms between \sigsfr and \sigstm, color-coded by the average \ubar, \qpah, and \sigd value of all the points that fall in that bin. 
The bins that are outside of the completeness bounds are shown with increased transparency and are not included in our fits. 
We fit a two dimensional plane of the form $Y = a~\log_{10}(\Sigma_\star) + b~\log_{10}(\Sigma_{\rm SFR})+ c$ and the results of these fits are listed in Table~\ref{TabCoeffs2d}.
In all cases, using two variables improves the resulting scatter compared to the single parameter fits listed in Table~\ref{TabCoeffsFitsResolved}.

In the left panels, we see that the prediction of \ubar using both \sigsfr and \sigstm results in RMSE of 0.2 dex, improving by 0.05 dex from the best single parameter predictions (with \sigsfr alone). The diagonal color gradient shows that both \sigsfr and \sigstm are correlated with \ubar.  

In the center column of Figure~\ref{Fig2DFits}, the top row shows a similar diagonal color gradient, implying that both \sigstm and \sigsfr also affect the PAH fraction; however, given the coefficients of the 2D fit, \sigsfr matters more than \sigstm. Additionally, in this case, the contribution of \sigstm appears to be positive, as opposed to being slightly negative in the one-dimensional fit. This is likely the result of the \sigstm--\sigsfr information allowing separation of radial trends in these quantities.
In general, the scatter in the predicted \qpah values are only improved a small amount by the two dimensional fits. For instance, as listed in Table~\ref{TabCoeffsFitsResolved}, \dmsr alone can predict \qpah to RMSE of 1.39\%. The best two parameter fit (\sigstm and \sigsfr) improves this slightly to RMSE of 1.2\%.

We perform the same 2D-fits to predict the dust surface density. In the one parameter fits, \sigsfr provided the best RMSE of 0.31 dex.  The two parameter fits improve this slightly to 0.26~dex for the combination of \sigstm and \sigsfr.  It is worth noting that these fits show that it is possible to predict the \sigd to within less than a factor of 2 scatter using only information about \sigstm and \sigsfr.

\input{tab_planefits}

\section{Discussion: Dust properties across the local Universe}
\label{SecDustProperties}

\subsection{Drivers of the Radiation Field Distribution in Galaxies}
\label{sec:disc_rad}
The radiation field intensity distribution in galaxies is of fundamental importance to understanding their dust emission. Radiation field intensity also plays critical roles in the physics of the ISM (regulating phase balance, H$_2$ formation, etc). At the scales we resolve in nearby galaxies (hundreds of pc to kpc), there is inevitably a distribution of radiation field intensities heating the dust. Knowledge of this radiation field distribution is critical for extracting information about changes in dust properties from the SED. For instance,  single-temperature modified blackbody models that fit an emissivity slope $\beta$ are subject to biases if the peak of the thermal SED is broadened by a radiation field distribution \citep{Hunt2015}. In addition, measuring changes in ``very small grain'' abundance is often  degenerate with the assumed radiation field distribution \citep{Galliano2022}.

Inferring the properties of this radiation field distribution can be approached in several ways. At the most detailed level, radiative transfer modeling of galaxies can attempt a full solution at some spatial resolution \citep[e.g.,][]{DeLooze2014, Nersesian2020}. However, radiative transfer modeling with sufficiently high resolution and realistic dust, SFR, and \stm distributions remains computationally challenging. Radiative transfer models for ``pieces'' of galaxies \citep{Law2018} are an alternative that yields insights into the radiation field distribution. A less computationally-intensive approach that provides information on radiation field distributions is ``energy-balance'' SED fitting, using codes like CIGALE \citep{Noll2009, Boquien19}, Prospector-$\alpha$ \citep{Leja2017}, or MAGPHYS \citep{daCunha2010}. These codes use assumptions about how attenuated starlight from various stellar populations translates to the re-radiated IR SED.  Fitting the UV to IR SED with these models provides constraints on the radiation field distributions, given those assumptions.  Such modeling has been widely used for unresolved galaxies \citep[e.g.,][]{Hunt2019, Nersesian19} and has been applied to small sets of resolved galaxies as well \citep[e.g.,][]{Boquien2012, Decleir2019, Abdorrouf2022}, though the spatial scales on which dust energy balance assumptions break down have not been fully quantified. 

Another approach, and the one we use in this work, is to assume an empirically-motivated shape for the radiation field distribution and constrain its parameters based on comparison to observed mid- to far-IR SEDs. This is possible because, having fixed the grain population properties, the radiation field intensity sets the equilibrium dust temperature {\em and} determines the intensity of the mid-IR stochastic dust emission (which is $\sim$~linear with $U$). A distribution of radiation field intensities will therefore broaden the thermal peak of the SED while maintaining the linear proportionality of the dust-mass weighted \ubar and the stochastic emission. We have adopted a simple model to describe the radiation field distribution, following \citet{DL07}, with a power-law plus delta function. The delta function is intended to describe dust in the diffuse ISM heated by the average interstellar radiation field.  The power-law distribution is intended to describe the higher intensity radiation fields in the vicinity of star forming regions. While we do not have information about the true distribution of radiation fields in our galaxies to compare against, we can compare the resulting correlations of the radiation field parameters with environment to assess their plausibility.

In Sections~\ref{sec:rad_int} and~\ref{sec:rad_res} we described the integrated and resolved correlations of radiation field distribution with galaxy properties.  We found that overall, the average radiation field intensity, \ubar, was similar to \umin, and the fraction of the dust luminosity from the power-law distribution, $\gamma$, was small.  This suggests that the majority of dust heating is due to the average interstellar radiation field, not to the very high intensity regions near current star formation. We also investigated how \ubar\ was correlated with galaxy properties. From our resolved analysis, we found that \ubar depends on both \sigsfr and \sigstm, suggesting both current star formation and older stellar populations contribute to the average radiation field intensity. These results are in good agreement with radiative transfer studies, which suggest that older stellar populations play a significant role \citep[$\sim$~tens of \%;][]{DeLooze2014,Viaene2017,Nersesian2020,Law2021,Rushton2022} in dust heating. This is also the case for results of UV--IR energy balance SED modeling (e.g., CIGALE), which find important contributions from older stellar populations to dust heating \citep{Nersesian19}. 

One clear trend from our analysis is the lack of correlation of $\gamma$ (the fraction of the dust luminosity from the power-law component) with quantities describing the stellar mass (\stm or \sigstm) and a strong correlation with parameters describing star formation (\sigsfr, \ssfrr, \dmsr). This can be seen in Figures~\ref{FigIntStarU} and~\ref{FigStarU}. This behavior for the power-law component is expected, and suggests the SED modeling is doing a reasonable job capturing the high intensity parts of the radiation field.

In total, the average radiation field, \ubar, in a $\sim$~kpc region of a galaxy has contributions from recent star formation both through $\gamma$ and through \umin. In Section~\ref{sec:planefits}, we showed that plane fits to a combination of \sigsfr and \sigstm allows \ubar to be predicted to $\sim0.2$~dex accuracy. 
This behavior suggests that dust heating from young stellar populations is strongly correlated with sSFR. These observations agree with the results of radiative transfer models, which also find a strong correlation between the young stellar heating fraction and sSFR \citep[e.g.,][]{Nersesian2020}, and with the results of energy balance SED fitting \citep[][]{BoquienSalim2021}, which suggest a stronger trend between the dust temperature (here equivalent to \ubar) and the sSFR, rather than SFR or \stm alone. A strong sSFR dependence is also consistent with analytical modeling results from \citet[][]{Hirashita22} and \citet[][]{Chiang23}.
The correlation of dust temperature (and \ubar) with sSFR on galaxy integrated scales is a topic of much interest in studies of high redshift galaxies, a topic we will revisit in Section~\ref{sec:highz}. 

\subsection{The Dependence of PAH Fraction on Integrated and Resolved Galaxy Properties}\label{SecPAHFraction}
PAHs are a critical component of the ISM, participating in many processes like photoelectric heating, H$_2$ formation, and UV attenuation \citep{Tielens08}.  However, fundamental aspects of their life cycle remain mysterious. Two key observations that provide insights into the PAH life cycle include: 1) their abundance relative to dust mass strongly depends on metallicity \citep{Engelbracht2005, Madden06, Draine07, Engelbracht2008, Gordon2008, Sandstrom2010, RemyRuyer15, Aniano20, Shivaei2024, Whitcomb2024}, and 2) they are destroyed in \ion{H}{2}~regions \citep{Cesarsky1996, Povich2007, Chastenet19, Chastenet2023, Egorov2023, Sutter2024}. 

Observations of the Magellanic Clouds provide particularly clear evidence for both of these trends. \citet{Chastenet19} found that at the LMC's metallicity of 1/2 Z$_{\odot}$, the diffuse neutral gas had PAH fractions similar to the Milky Way diffuse neutral gas \citep[$\sim4-5$\%;][]{DL07}, while \ion{H}{2}~regions showed steep drops in PAH fraction.  Moving to the SMC, at 1/5 Z$_{\odot}$, the PAH fraction in the diffuse ISM was far lower ($\sim1$\%), with \ion{H}{2}~regions showing yet lower values.  These observations suggest a steep metallicity dependence in PAH fraction that sets in below $\sim1/2$~Z$_{\odot}$. Recent work by \citet{Whitcomb2024} measured \qpah as a function of radius in M101, NGC~628, and NGC~2403, revealing a steep drop in at a threshold metallicity of 2/3 Z$_{\odot}$. With the mid-IR resolution of JWST, it is now possible to separate \ion{H}{2}~regions from the diffuse ISM in many nearby galaxies and examine trends in mid-IR traced PAH fraction both in and out of \ion{H}{2}~regions.  \citet{Sutter2024} present such observations for 19 nearby galaxies, showing an approximately constant PAH fraction outside \ion{H}{2}~regions and steep drops within them at $\sim$~Z$_{\odot}$. They demonstrate that $\gtrsim$~kpc-scale PAH fraction in $\sim$~Z$_{\odot}$ galaxies is highly correlated with sSFR, since it traces a relative fraction of the ISM that is currently in \ion{H}{2}~regions.  

Our observations provide a comprehensive view of the $\sim$~kpc scale and integrated \qpah in nearby galaxies and allow us to control for completeness in measuring the scaling of \qpah with environment.  One of our key findings is that \qpah is a steeply decreasing function of sSFR and \ssfrr (Figures~\ref{FigIntStarDust} and~\ref{FigStarDust}): \qpah is strongly, negatively, correlated on resolved scales with all quantities that trace the local intensity of star formation (\sigsfr, \ssfrr, and \dmsr).  On galaxy-integrated scales, strong negative correlations with sSFR and \dms persist, though the \qpah does not vary strongly with galaxy-averaged SFR. It is clear from this summary that the local intensity of star formation is a primary agent in setting the PAH fraction in galaxies, likely due to the destruction of PAHs in \ion{H}{2}~regions.

Strong trends in PAH fraction correlated with sSFR have also been seen in integrated galaxy measurements \citep{RemyRuyer15,Nersesian19,Galliano2021} and resolved studies \citep[e.g., the trends with 70 \micron\ surface brightness in][]{Aniano20}.  Using CIGALE \citep[][]{Boquien19}, and The Heterogeneous dust Evolution Model for Interstellar Solids \citep[THEMIS;][]{Jones17}, \citet[][]{Nersesian19} fit integrated UV-to-far-IR SEDs of the DustPedia sample \citep[][]{Davies17, Clark2018}. They studied the variations of the small carbonaceous grain content as a function of galaxy morphology and found that the fraction of small grains decreases with sSFR after a threshold ${\rm log_{10}(sSFR) \geq -10.5}$, in good agreement with our findings. They also suggest that this decrease of the small grain fraction is due to the harder radiation fields found at higher sSFR.
\citet[][]{RemyRuyer15} found similar trends using the KINGFISH and DGS \citep[Dwarf Galaxy Survey;][]{Madden13DGS} galaxy samples.
Our findings align very well with the previously observed scaling relations, but now extend them to resolved scales across the galaxy population.

Because of the shape of the star forming main sequence, sSFR is higher for lower \stm galaxies (and \ssfrr is higher for lower \sigstm).  To attempt to separate out \stm trends \citep[which may be tied to metallicity via the mass-metallicity relationship;][]{tremonti2004}, we investigated the correlation of PAH fraction with offset from the main sequence (\dms and \dmsr). The observed trends of \qpah with \dms and \dmsr are shallower than those with sSFR and \ssfrr, suggesting that sSFR trends may include a combination of both metallicity and sSFR effects. However, we also observe a slightly negative slope of \qpah as a function of \sigstm. The observed dependence of \qpah on metallicity, plus the generally increasing metallicity towards the centers of galaxies would lead us to expect a positive trend between \qpah and \sigstm. Our observed trend, which we note does have significant scatter, may be the result of the higher \sigsfr and \ssfrr towards galaxy centers dominating over the gradient of metallicity and \sigstm in setting \qpah.  Another option may be the contribution of softer radiation fields to heating PAHs at high \sigstm. Such an effect has been found to alter \qpah derived from mid- to far-IR modeling in the center of M31 \citep{Draine2014} and in other galaxy centers \citep{Whitcomb2024}.

It is worth noting that in many studies of the metallicity dependence of the PAH fraction, galaxy samples have inevitably been limited by the need to find bright, low metallicity targets. Selecting low metallicity galaxies with detectable mid-IR emission has favored galaxies with unusually high star formation rates for their masses (e.g., ``blue compact dwarfs'' like I~Zw~18, Henize 2-10, Haro~11, and others).  Given the strong dependence of PAH fraction on sSFR and \dms, focusing on these highly star forming galaxies may lead to overestimation of metallicity effects.  This is also true for targeted spectroscopic studies of individual star forming regions in galaxies---without sampling ``average'' star formation intensities, these observations may lead to biased conclusions about the strength of the metallicity dependence of the PAH fraction. We do not have detailed metallicity information for most galaxies in our sample and our sample is not complete down to very low metallicity (stellar mass), so we cannot draw strong conclusions about metallicity dependence for \qpah. Our strong trends of \qpah with \dms and \dmsr show that there are large variations in \qpah even controlling for metallicity.  Separating sSFR and metallicity effects therefore requires care towards completeness and unbiased samples of galaxies. Future studies with JWST, given its high sensitivity, should be able to disentangle the dependence of PAH fraction on sSFR and metallicity, given careful sample selection.

\subsection{The Relationship between Dust Mass, Stellar Mass, and Star Formation Rate}
\label{SecDustandStellar}
Trends in \mdust with SFR and \stm likely reflect the combined effects of several well-known scaling relationships: 1) the mass-metallicity relationship or fundamental mass-metallicity-SFR relationship \citep{Maiolino2019}, which shows that lower \stm galaxies tend to have lower metallicity; 2) the Kennicutt-Schmidt relationship \citep{KennicuttEvans12KSReview}, which describes the scaling between gas mass and SFR; 3) the gas mass to stellar mass scaling \citep{Saintonge2022}, which describes the behavior of gas fraction over the star-forming main sequence; and 4) the dependence of the dust-to-metals ratio on metallicity \citep[which sets the mapping between metallicity and dust-to-gas ratio;][]{remy-ruyer2014,DeVis2019,Peroux2020}. From the combination of these scaling relationships, we expect predictable trends in \mdust to arise over the SFR--\stm space. Such trends have been explored by a number of studies for integrated measurements of galaxies \citep{Cortese2012, DeVis17a, Casasola2020, DeLooze2020}. These scaling relationships also have resolved behavior \citep{Leroy08, Bigiel08, Sanchez2014, Hsieh2017, Ellison2018, Lin2019}, which should similarly result in correlations of \sigd with \sigsfr, \sigstm, \ssfrr, and \dmsr. Our observations provide a powerful database with which to explore these trends, carefully controlling for completeness in \sigd and selecting where we should be able to measure accurate SFRs.

\begin{figure*}
    \centering
    \includegraphics[width=\textwidth]{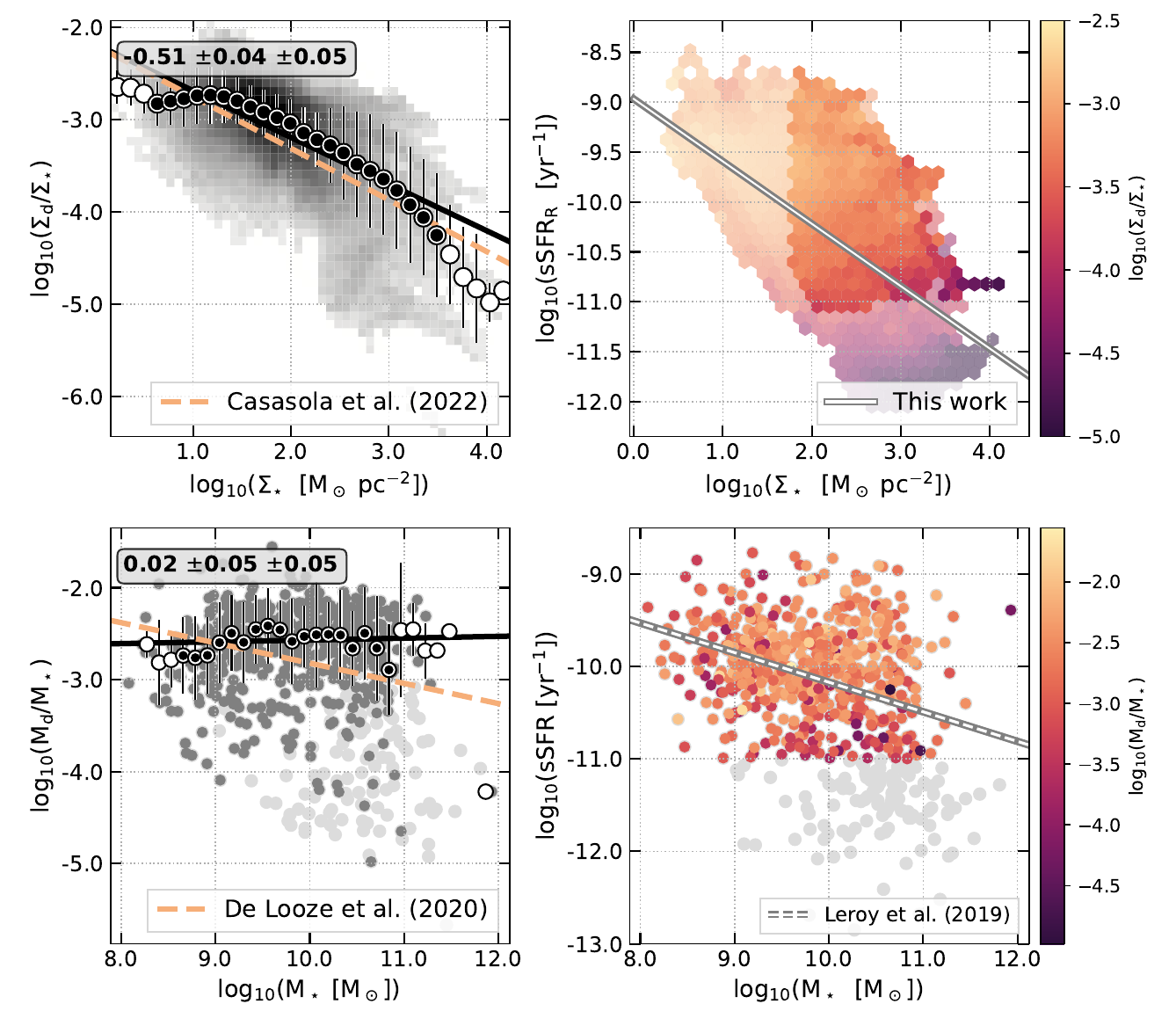}
    \caption{
    \textit{Left column:} specific dust mass as a function of stellar mass on resolved (top) and integrated (bottom) scales. The light grey points are those with $\rm \log_{10}(sSFR)<-11$. The medians are measured on the dark grey points. error bars show the scatter within the bin; the solid black line is the best-fit using the filled medians, whose slope is reported in the label; the first uncertainty is the statistical error on the fits and the second uncertainty is the standard deviation on coefficients when fitting binned data with a varying number of bins.
    Orange dash lines show best fit from the literature: best-fit from \citet{Casasola2022} for a few galaxies in the DustPedia sample, and from \citet{DeLooze2020} for a large sample of integrated measurements.
    \textit{Right column:} main-sequence color-coded by the specific dust mass. In the resolved panel (top), the transparency shows all pixels, while the opaque bins cut out pixels with $\log_{10}(sSFR) < -11$ and above the completeness threshold in \sigstm. The line shows our main-sequence fit from Section~\ref{sec:surfacedensities}. 
    Bottom panel shows the integrated values from this work for the main-sequence, and the white-dashed line shows the main-sequence fit by \citet{Leroy19}.
    }
    \label{FigSpecificDustStar}
\end{figure*}

On integrated scales, above our sSFR cut ($\rm \log_{10}(sSFR)<-11$), we find a remarkably constant \mdust/\stm ratio spanning $\log_{10}(\rm M_\star)=9-11$ (indicated by the slope of $\sim1$ in Figure~\ref{FigIntStarDust} and the nearly constant ratio in the bottom-left panel of Figure~\ref{FigSpecificDustStar}).  A variety of other integrated studies of galaxies in the local universe have drawn  different conclusions, finding negative trends of \mdust/\stm with increasing \stm \citep{Cortese2012,DeVis17a,Casasola2020,DeLooze2020}. \citet{DeLooze2020} found a slope of $-0.22$ fitting the integrated SEDs of 423 galaxies, a combination of the JINGLE \citep{Saintonge2018}, KINGFISH \citep{Kennicutt11KINGFISH}, HAPLESS \citep{Clark2015HAPLESS}, HiGH \citep{DeVis17a} and some of the HRS \citep{Boselli2010} samples. We can reproduce negative slopes by relaxing our sSFR cut, as can be seen in Figure~\ref{FigSpecificDustStar}. Here the light grey points show galaxies that were excluded by our sSFR cut---they systematically fall below the average \mdust/\stm. Including these points in our fit would yield a negative slope for \mdust/\stm versus \stm. The primary driver of this trend is likely to be the decreased gas-to-\stm fraction in these low sSFR galaxies \citep{Cortese2012,Saintonge2022}.  

Because of the mass-metallicity relationship, lower~\stm galaxies are expected to have lower metallicity, and hence lower DGR.  In addition, lower mass galaxies are observed to have higher gas fractions and higher sSFR.  These DGR and gas fraction trends should therefore compete with each other in setting the \mdust/\stm scaling as a function of \stm.  In Figure~\ref{FigSpecificDustStar}, we show in the bottom right panel, the sSFR--\stm space color coded by \mdust/\stm. The figure shows that on or near the star forming main sequence the trends at least approximately cancel out, leaving a $\sim$~fixed \mdust/\stm ratio over $>2$ orders of magnitude in \stm. 

Distinct from the constant \mdust/\stm ratio we see for integrated galaxies, on resolved scales \sigd/\sigstm is a decreasing function of \sigstm, with slope $-0.51$, as shown in Figure~\ref{FigSpecificDustStar} in the top-left panel.  This trend is very similar to the slope of $-0.56$ observed in a sample of 18 face-on spiral galaxies from DustPedia \citep{Casasola2022} and also resembles high spatial resolution measurements of the \sigd--\sigstm scaling in M31 \citep{Viaene2014}. This measurement indicates that there is a sub-linear power-law slope for the \sigd--\sigstm relationship (also shown in Figure~\ref{FigStarDust}). Furthermore, because the majority of galaxies have negative metallicity gradients, a given \sigd value drawn from the inner or central part of a galaxy will tend to correspond to less $\Sigma_{\rm gas}$ than the same \sigd in the outskirts of the galaxy. This implies that most galaxies will also have a sub-linear $\Sigma_{\rm gas}$--\sigstm relationship. 

In outer disks, where \ion{H}{1} represents most of the ISM mass, such a sub-linear $\Sigma_{\rm gas}$--\sigstm\ relationship is in good agreement with the frequent observation that the atomic gas disks of galaxies often extend to larger radii compared to the stellar disks, with the $\Sigma_{\rm gas}/$\sigstm ratio increasing with increasing galactocentric radius \citep[e.g.,][]{Bigiel2010,KennicuttEvans12KSReview,Wang2016}. However, at \sigstm$\gtrsim100$~\msun~pc$^{-2}$, where the ISM is H$_2$-dominated \citep{Leroy08,Schruba2011}, many observations show a $\sim$~linear power-law relationship between CO emission and \sigstm \citep{Leroy08,Schruba2011,Lin2019,Sanchez2020,Pessa2022} suggesting that a linear relationship between $\Sigma_{\rm mol}$--\sigstm. This difference in the CO--\sigstm and \sigd--\sigstm slopes may be the result of systematic variations in the CO-to-H$_2$ conversion factor ($\alpha_{\rm CO}$). Recent work by \citet{Chiang2024} uses measurements of CO, \ion{H}{1}, \sigd, and metallicity for a sample of 37 nearby galaxies, with an assumption of a constant dust-to-metals ratio, to infer $\alpha_{\rm CO}$ that decreases with increasing \sigstm, $\alpha_{\rm CO}^{\rm 2-1} \propto \Sigma_*^{-0.5}$ (for CO J$=$2$-$1, more shallowly for CO 1$-$0). From our compilation of \ngres resolved galaxies, we can see that the shallow \sigd--\sigstm is a general feature of the sample. This observation, combined with the widely observed, linear CO--\sigstm relationship suggests that a decreasing $\alpha_{\rm CO}$ with \sigstm is a general trend across the galaxy population.

As in the integrated case, we find that \sigd/\sigstm is positively correlated with the offset above the resolved main sequence (\dmsr), as shown in Figure~\ref{FigSpecificDustStar} in the top right panel. Assuming that fixed \sigstm controls for metallicity and DGR variations to first order, this trend would suggest that points are offset above the resolved main sequence (high \dmsr) because they have a larger dust (and gas) surface density.  This result agrees with recent work showing that the correlation properties of $\Sigma_{\rm mol}$--\sigsfr--\sigstm suggests that offset from the resolved main sequence is primarily driven by having higher $\Sigma_{\rm mol}$ \citep{Baker2022}. We note, however, that conclusions drawn about offsets in $\Sigma_{\rm mol}$ are subject to uncertainties in $\alpha_{\rm CO}$, while \sigd may have a less variable relationship to gas.

\subsubsection{The \sigd--\sigsfr Relationship}\label{sec:inverseks}
The \sigd--\sigsfr relationship is closely tied to the resolved star formation scaling law or Kennicutt-Schmidt relationship, which relates gas surface density ($\Sigma_{\rm gas}$) and \sigsfr \citep{KennicuttEvans12KSReview}, as well as the closely related concept of the local gas depletion time, $\tau_{\rm dep} = \Sigma_{\rm gas}/$\sigsfr . There has been substantial effort to characterizing this relationship. The slope is found to vary depending on whether the ISM is \ion{H}{1}- or H$_2$-dominated \citep{Leroy08,Bigiel08}. Shorter gas depletion times and more nearly linear slopes are often found in regions of galaxies dominated by molecular gas, while the \ion{H}{1}-dominated parts of galaxies show a much wider range of depletion times, often including very long $\tau_{\rm dep}$ in outer galaxies, and steeper $\Sigma_{\rm SFR}$--$\Sigma_{\rm gas}$ relationships. This is understood to reflect that stars form from molecular gas, and that the balance of phases in the ISM is tied to local environmental conditions, including dynamical equilibrium pressure in the ISM and the dust-to-gas ratio \citep{Blitz2006, Leroy08, Sun2020, Eibensteiner2024}. 

The measured slope in the H$_2$-dominated ISM is sensitive to the assumed behavior of the CO-to-H$_2$ conversion factor \citep{Sun2023, Teng2024}. However, many measurements of the CO-to-H$_2$ conversion factor ($\alpha_{\rm CO}$) are based on \sigd, with built-in assumptions about the dust-to-gas ratio (DGR). This makes all aspects of this problem closely connected \citep{Chiang2024}. For high redshift galaxies, the \sigd--\sigsfr scaling may be more observationally accessible than the relationship between \ion{H}{1} or H$_2$ and \sigsfr, making our observed relation useful in its own right. Since the ratio $\Sigma_{\rm d}/\Sigma_{\rm SFR}$ may be even more accessible at high-$z$ than the scaling, we note that for an intermediate $\Sigma_{\rm d} = 0.1$~M$_\odot$~pc$^{-2}$, our fit implies a ``dust depletion time'' 
\begin{equation}
\label{eq:taudust}
\tau_{\rm dep}^{\rm dust} \equiv \frac{\Sigma_{\rm dust}}{\Sigma_{\rm SFR}} \approx 8.7~{\rm Myr}~.
\end{equation}
Though star formation is clearly fueled by gas, not only dust, this ratio expressed the star formation rate per unit dust and may be an interesting point of comparison for high-$z$ systems or other populations.

\begin{figure}
    \centering
    \includegraphics[width=0.5\textwidth]{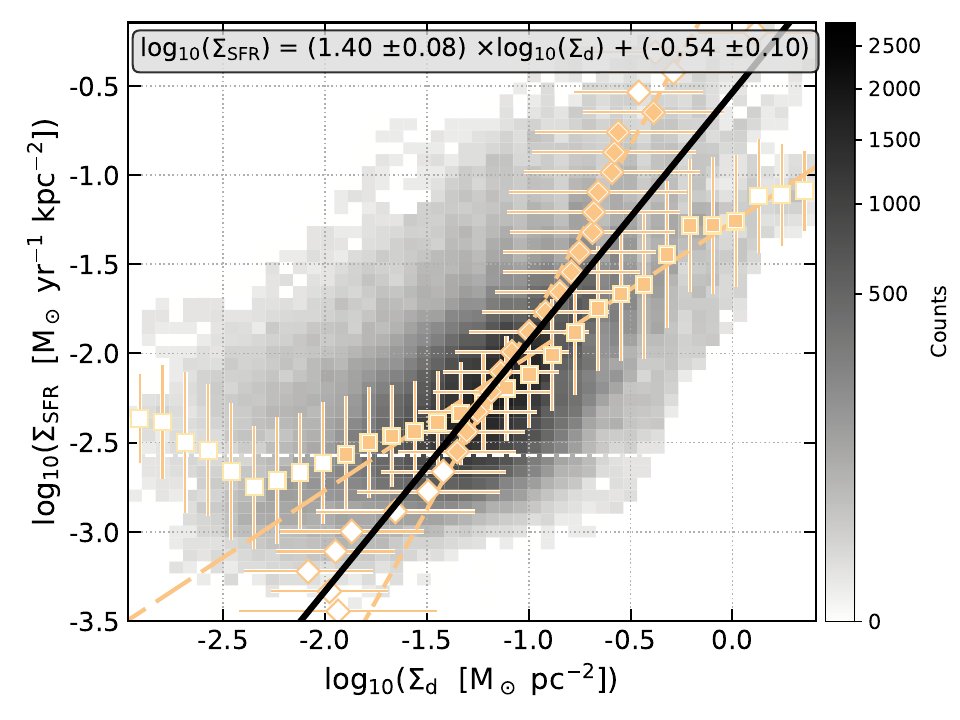}
    \caption{2D histogram and scaling relation of the dust surface density, \sigd, and the star formation rate surface density, \sigsfr. The orange medians and lines are derived from binning either in $x$ or $y$, and the black line shows the mean of these slopes, with the reported equation.}
    \label{FigDustKS}
\end{figure}

As has been extensively discussed in the literature \citep{Blanc2009,Leroy2013,Shetty2014}, methodological choices in binning and power-law fitting play an important role in determining the resulting slope of the star formation scaling relation. 
In Figure~\ref{FigDustKS}, we show slopes for fits to the binned averages, with binning set either in $x$ or $y$, in light orange.
Because of the intrinsic scatter in the data, binning by one or the other axis changes the slope and biases the derived scaling relation. To minimize these biases related to the choice of the dependent variable, we use the ``bisector'' fit (in black), which takes the mean of the two slopes. 
The fit is described by the equation:   
\begin{equation}
    \log_{10}(\Sigma_{\rm SFR}) = (1.40 \pm 0.08) \times \log_{10}(\Sigma_{\rm d}) - (0.54 \pm 0.10),
\end{equation}
with \sigsfr in \msolkpcsqyr and \sigd in \msolpcsq.  The slope we fit is in good agreement with recent work by \citet{Chiang23}, who also derived a \sigsfr--\sigd relationship.

Our observed \sigd--\sigsfr fit, assuming a constant DGR, provides a measurement of the \sigsfr--$\Sigma_{\rm gas}$ scaling law.  
Our observed slope of 1.4 is slightly steeper than the slope of $1.2$ obtained for only molecular by \citet{Sun2023} using dust-based $\alpha_{\rm CO}$ estimates (their B13 values) or similar results by \citet{Teng2024}. This likely reflects that our measurements include dust and SF associated with both the inner H$_2$-dominated and the outer \ion{H}{1}-dominated parts of disks.  At ${\rm DGR=0.01}$, typical for Solar metallicity ISM, the lowest surface densities included in our fit, $\log_{10}(\Sigma_{\rm d}~[{\rm M_\odot~pc^{-2}}]) \sim -1.5$, would correspond to a total gas mass surface density as low as $\Sigma_{\rm gas} \sim 3$~\msun~pc$^{-2}$. This gas surface density range extends well into the \ion{H}{1}-dominated parts of disks, showing that we are not only sampling the more linear \sigsfr--$\Sigma_{\rm gas}$ relationship in H$_2$ dominated regions. This appears to be in good agreement with measurements of the $\Sigma_{\rm SFR}$--$\Sigma_{\rm gas}$ scaling relation that include both \ion{H}{1} and H$_2$ \citep[][]{Bigiel08, Schruba2011, KennicuttEvans12KSReview}.

Alternatively to fitting the \sigd--\sigsfr slope, we could assume a gas depletion time to convert the observed \sigsfr into a gas mass surface density, from which we could determine the DGR \citep[i.e., ``inverting the Kennicutt-Schmidt relation'', e.g.,][]{Genzel2012,Schruba2012}. One option would be to estimate a $\tau_{\rm dep}$ that depends on $\Sigma_{\rm gas}$ according to the galaxy-integrated $\Sigma_{\rm gas}$--\sigsfr relation from \citet{Kennicutt98}, which has a power-law slope of 1.4 and intercept of $2.5 \times 10^{-4}~{\rm M_\odot~yr^{-1}~kpc^{-2}}$. In this case, converting $\Sigma_{\rm SFR}$ to $\Sigma_{\rm gas}$ and comparing to $\Sigma_{\rm dust}$, we estimates a  median DGR of ${\rm log_{10}(DGR_{KS}) = -2.2\pm0.3}$ across our whole data set. This value is slightly lower, but relatively consistent with the range of ${\rm M_d/M_H}$ ratios found in Milky Way (i.e., Solar metallicity) dust models: 0.01 in the \citet{Compiegne2011} model, 0.0074 in THEMIS \citep{Jones17}, or 0.01 in \citet{Hensley2023}.
Additionally, we make the same measurement with the $\Sigma_{\rm H_2}$--\sigsfr relationship from \citet{Sun2023}, appropriate for the H$_2$ dominated ISM. We use their `mKS' prescription and find ${\rm log_{10}(DGR_{mKS}) = -2.4\pm0.4}$. We note that \citet{Sun2023} adopt a metallicity-varying CO-to-H$_2$ conversion factor, which is calibrated using dust as a tracer of total gas \citep{Bolatto2013}. In fact, it is worth noting that observations of DGR, the CO-to-H$_2$ conversion factor, and molecular gas depletion time ($\Sigma_{\rm mol}$/\sigsfr) are all observationally connected through the uncertainty in converting CO to H$_2$. 

\subsection{\qpah and WISE bands}
Several studies have focused on using band ratios of mid-IR photometry as tracer of PAH fraction, often contrasting a PAH-dominated band with a continuum-dominated one. This was done with \textit{Spitzer} at $\sim 8$ and 24~$\mu$m \citep[e.g.,][]{Engelbracht2005, Smith2007, Engelbracht2008, Marble2010, Croxall2012}, and more recently with JWST \citep[e.g.,][]{Chastenet2023, Egorov2023, Sutter2024}.
Although they do not cover exactly the same wavelengths, the WISE~3 ($\sim12$~\micron) and WISE~4 ($\sim22$~\micron) bands can be used for similar purposes.
Additionally, because WISE performed an all-sky survey, a PAH fraction tracer constructed from these two bands would be available for all WISE-detected galaxies.
To facilitate this, we examine the relation between WISE~3/WISE~4 ratios and the model parameter \qpah.

\begin{figure}
    \centering
    \includegraphics[width=0.5\textwidth]{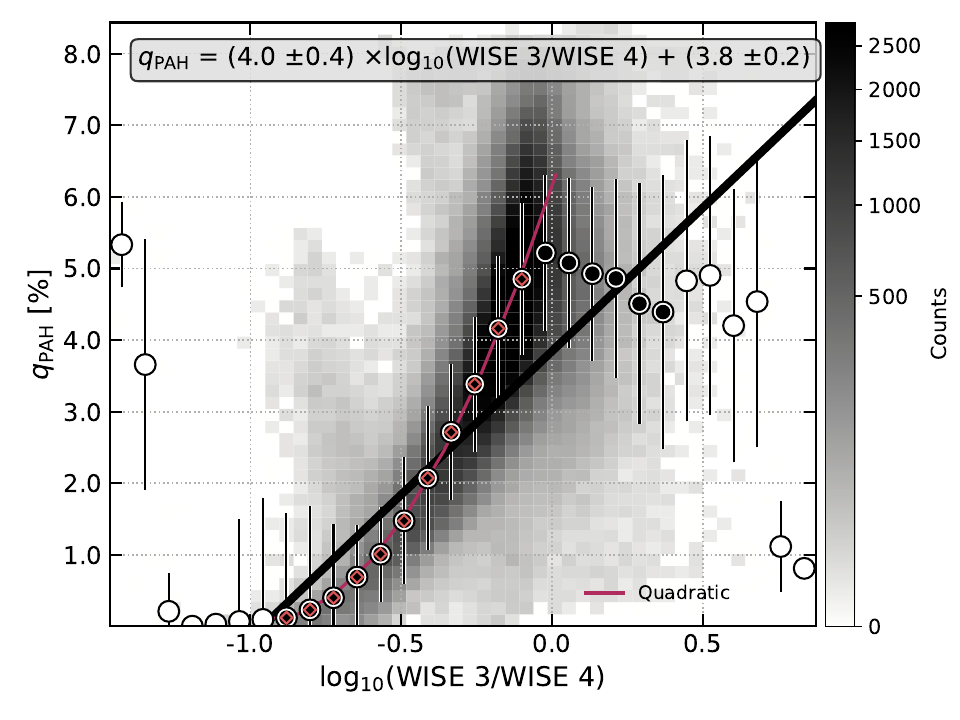}
    \caption{Fitted \qpah from the resolved analysis as a function of ${\rm \log_{10}(WISE~3/WISE~4})$ ratio. We fit a line through the bins that pass the same cuts used in other figures, for consistency, marked in black-filled symbols. 
    For prediction purposes only, we show an additional fit in the form of a quadratic fit (see text).}
    \label{FigWISEqPAH}
\end{figure}

In Figure~\ref{FigWISEqPAH} we show the relation between the ratio of WISE~3-to-WISE~4 emission and the value of \qpah in all the fitted pixels of our sample (on resolved scales, above the 3$\sigma$ detection). The white circles are the median values of \qpah in bins of \logt(WISE~3/WISE~4).

For consistency with the rest of the analysis, we perform linear fits to these medians, selecting only those with at least 250 points per bin, and show the result of that fit in the figure. This fit is described by the following equation:
\begin{equation}
\begin{split}
    q_{\rm PAH} = (4.0 \pm 0.4) \times \log_{10}({\rm WISE~3/WISE~4}) + \\
    (3.8 \pm 0.2).
\end{split}
\end{equation}
However, there is clearly variation not consistent with a simple linear function visible in the binned medians.

Going further, for the sake of providing a predictive functional form only (without physical interpretation), we fit a quadratic function, shown in solid red line, to that same distribution of medians. These are marked with red diamonds. 
For the purpose of estimating \qpah with WISE bands, for values of the WISE~3/WISE~4 ratio between 0.09 and 0.9, which covers the majority of the dataset, we suggest
\begin{equation}
\begin{split}
    q_{\rm PAH} = &(6.6 \pm 0.2) \times \log_{10}({\rm WISE~3/WISE~4})^2 + \\
            &(12.7 \pm 0.3)\times \log_{10}({\rm WISE~3/WISE~4}) \\
            &(6.2 \pm 0.1).
\end{split}
\end{equation}

\subsection{Applications to High Redshift Studies}\label{sec:highz}
Observations of distant galaxies are generally limited in their spatial resolution and wavelength coverage for studying dust emission.  At long wavelengths, many studies rely on $\sim2$ rest-frame far-IR measurements to fit an SED \citep[e.g.,][]{Witstock2023} to constrain both dust mass and temperature for a galaxy. At shorter wavelengths, JWST has enabled sensitive observations of mid-IR emission from galaxies out to redshifts of $z\sim3-4$ \citep{Ronayne2023, Shivaei2024, Alberts2024}, but does not cover the $\sim20$ \micron\ continuum emission, leaving diagnostics of PAH fraction uncertain.  For measuring dust mass and PAH content of high-$z$ galaxies from their infrared emission, the key missing information is the intensity and distribution of the radiation field. Since many high-$z$ galaxies have measurements of \stm and SFR, and often \dms by comparison to the galaxy population at that redshift, it may be possible to use our scaling relationships to infer the radiation field properties for these galaxies. A known radiation field distribution would enable more accurate estimates of \mdust and \qpah from limited mid- or far-IR SED coverage. 

\begin{figure*}[ht!]
    \centering
    \includegraphics[width=\textwidth]{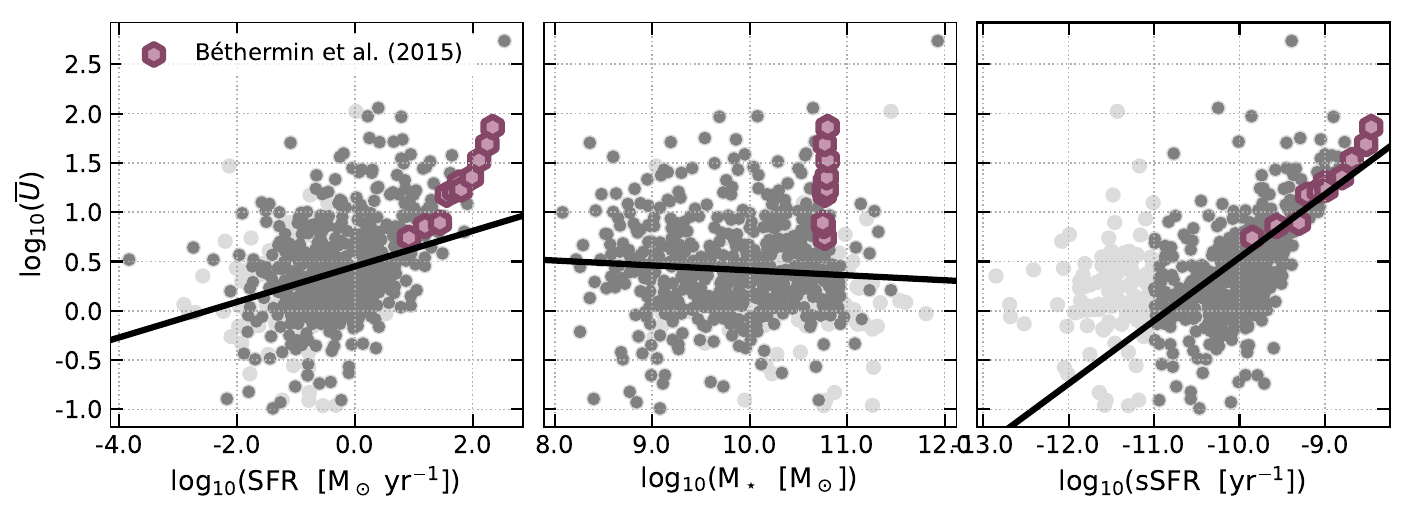}
    \caption{Distribution of SFR, \stm, and sSFR as a function of the integrated fit value of \ubar, for the sample in this paper (grey points), with the best-fit linear relations shown as solid black lines (Figure~\ref{FigIntStarU} and Table~\ref{TabCoeffsFitsInt}). The purple symbols are from the main-sequence sample of \citet{Bethermin2015}.}
    \label{fig:z0MGvsBethermin}
\end{figure*}

A key question is whether it is possible to extrapolate the scaling relations we have determined for $z=0$ galaxies for use at higher redshifts. To test this, we would ideally need similar SED coverage for high-$z$ galaxies, in order to apply the same type of modeling and derive radiation field properties. While most individual galaxies only have sparse SED coverage, it is possible to stack galaxies to obtain average SED constraints.  \citet{Bethermin2015} used this approach for galaxies at $0<z<4$ from the COSMOS field using data from surveys of the field by \textit{Spitzer}, \textit{Herschel}, LABOCA, and AzTEC. They performed stacks for a sample of main sequence galaxies at each redshift, as well as a sample of starbursts, defined to be 10 times the sSFR of the main-sequence at that redshift.  With the stacked SEDs, they fit \citet{DL07} models, allowing them to obtain \umin, $\gamma$, and \ubar.  These results are directly comparable to our work. \citet{Bethermin2015} also provided average values of SFR, \stm, and sSFR for the stacked samples.  

To test our scaling relationships and their ability to predict radiation field for high-$z$ targets, in Figure~\ref{fig:z0MGvsBethermin} we show our integrated galaxy measurements of \ubar as a function of SFR, \stm, and sSFR along with the binned values for main-sequence galaxies from $0<z<4$ from \citet{Bethermin2015}.  We find an excellent correspondence between their measured \ubar and our predicted \ubar--sSFR scaling relationship, across redshift bins. Our SFR scaling does not provide a good prediction for \ubar, indicating that galaxy averaged SFR does not trace the conditions which change the radiation field distribution.  This makes sense given that a high SFR could result from either a large amount of distributed low intensity star formation, or a small, high intensity, region of star formation.  
\stm also fails at predicting \ubar for the high redshift samples, though this is also the case for the $z=0$ galaxies as well. 

The correlation of dust temperature and radiation field with sSFR has been highlighted in a variety of studies in addition to \citet{Bethermin2015}. \citet[][]{Magnelli2014} used stacked Herschel observations of galaxies in deep fields to measure average dust temperature in SFR-\stm space.  They found that dust temperature was more strongly correlated with sSFR and \dms than with SFR and \stm. \citet{Schreiber2018} also fit SEDs of stacked galaxies, reproducing the trend of dust temperature increasing with redshift and showing a decreasing trend of PAH fraction (traced by their IR8 parameter) as a function of redshift and metallicity.  \citet{Schreiber2018} also investigated trends relative to the main sequence and found correlations between the dust temperature and PAH fraction with the equivalent of \dms.  The cause of trends of dust temperature with sSFR or \dms have been explained through combinations of increased star formation efficiency (leading to higher SFR and radiation field intensity for a given \sigd), more compact star forming regions, or changes in the dust-to-gas ratio or dust-to-metals leading to more pervasive, higher intensity radiation fields for a given SFR \citep{Hirashita22,Sommovigo2022}.

The fact that local universe sSFR scaling does a good job of predicting \ubar up to $z\sim4$ is surprising, considering the very different properties of galaxy disks expected at that era. Galaxy disks and star forming clumps show high velocity dispersions and high \sigsfr along with disk scale heights that are thought to be much more vertically extended compared to $z\sim0$ galaxies \citep{Tacconi2020}. Given two galaxy disks with similar \ssfrr, we might expect that the disk with the larger scale height might have lower \ubar, simply by a geometric argument. However, galaxies at higher redshift may also have overall lower metallicity and dust-to-gas ratio, meaning that the same \sigsfr could produce a harder, more pervasive radiation field that can heat the available dust up to higher temperatures. 

\section{Conclusions}
\label{SecConclusions}
We present an analysis of dust and radiation field properties of a sample of more than 800 galaxies in the local universe based on WISE and \textit{Herschel} SED fitting. 
Using the \citet{DL07} dust model and the \dustbff grid-based Bayesian fitting code, we derive the properties of the distribution of radiation fields heating the dust (\umin, $\gamma$, \ubar), the dust mass surface density (\sigd), and the fraction of dust in the form of PAHs (\qpah) on both integrated and resolved scales. Our sample provides one of the largest, diverse samples of galaxies where resolved maps of \sigd and \qpah are available.

We fit scaling relations between the dust model parameters with SFR, \stm, sSFR, and offset from the main-sequence, \dms, on integrated scales (Figures~\ref{FigIntStarU}, \ref{FigIntStarDust}) and resolved scales (Figures~\ref{FigStarU}, \ref{FigStarDust}). Results of our analysis include:
\begin{enumerate}
    \item We find that all parameters describing the radiation field (\ubar, \umin, $\gamma$) are positively correlated with quantities related to star formation (SFR, sSFR, \dms) on both integrated and resolved scales. On resolved scales, the correlation of \umin and \ubar with both \sigstm and \sigsfr suggest that the average radiation field has contributions from both old and young stellar populations.
    \item The derived radiation field parameters show behavior consistent with expectations for the delta-function plus power-law model, where the high intensity radiation fields in the vicinity of star forming regions populate the power-law distribution and \umin describes the minimum interstellar radiation field.  We find $\gamma$ is generally small and shows weak or no correlation with stellar mass on either resolved or integrated scales, but strong correlation with SFR related quantities.
    \item We find a strong anti-correlation between the PAH fraction and sSFR and \dms on both integrated and resolved scales. For resolved \qpah, an increase of 1 dex in \ssfrr results in \qpah dropping from the $4-5$\% typical for solar metallicity ISM to $\sim1$\%, comparable to what is seen in low metallicity galaxies.  Resolved \qpah is also strongly anti-correlated with \sigsfr, emphasizing that the \textit{local} star formation is a key agent affecting the PAH fraction, likely due to destruction of PAHs in \ion{H}{2} regions. The strong anti-correlation of \qpah with sSFR suggests that careful sample selection is necessary to isolate trends of \qpah with metallicity, since many low-$Z$ targets have been selected to highly star-forming dwarfs and the shape of the SF main sequence already introduces a correlation of sSFR with \stm (and therefore metallicity).
    \item We find a constant ratio of \mdust/\stm with \stm on integrated scales for main sequence galaxies. Galaxies with low sSFR, falling below the main sequence, have lower \mdust/\stm, most likely due to low gas fractions. On the main sequence, we see a constant \mdust/\stm value over $>2$ orders of magnitude in \stm, suggesting that trends of metallicity (and DGR) and gas fraction with \stm cancel each other out in setting \mdust/\stm. 
    \item The constant \mdust/\stm does not persist on resolved scales, where we observe a negative correlation of \sigd/\sigstm with \sigstm. We find that \sigd/\sigstm increases at fixed \sigstm above the resolved main sequence.  This suggests that higher \sigd/\sigstm results from higher gas (and dust) surface densities, correlated with increased \sigsfr.  The overall sub-linear correlation of \sigd with \sigstm seems to be a general property of the distribution of the ISM in galaxies, and provides an interesting contrast to the widely observed linear CO-\sigstm correlation, highlighting systematic trends in $\alpha_{\rm CO}$.
    \item We find that \sigsfr$\propto \Sigma_{\rm d}^{1.4\pm0.08}$.  The slope of $\sim1.4$ suggest our \sigd measurements extend into the \ion{H}{1} dominated parts of galaxies.  We measure a typical ``dust depletion time'', analogous to the more standard gas or H$_2$ depletion time, of $\approx8.7$ Myr.  This value of SFR per unit dust mass may be a useful comparison for high-$z$ systems where gas masses, particularly atomic gas, may be more challenging to measure than dust masses.
    \item Using a combination of \sigsfr and \sigstm, our two-dimensional scaling relationships (Section~\ref{sec:planefits}) are able to predict \ubar to a root-mean square error of 0.2 dex, \qpah to 1.2\%, and \sigd to 0.26 dex.
    \item We assess the predictive power of our integrated-galaxy radiation field scaling relations for high-redshift galaxies, using stacked SEDs of main sequence galaxies from \citet{Bethermin2015}. We find that our \ubar--sSFR scaling relationship matches the \ubar values for the stacked main seqeunce galaxies from $0<z<4$.
\end{enumerate}

Associated with this paper, we deliver all reduced \herschel data and dust fitting results to the Infrared Science Archive (IRSA), as described in Appendix~\ref{AppDelivery} (doi: \href{https://www.ipac.caltech.edu/doi/10.26131/IRSA581}{{\tt 10.26131/IRSA581}}). We note that our delivery includes \herschel data for a substantial number of galaxies that were not included in our analysis, due to lacking both PACS and SPIRE coverage (or WISE data in the \zzmgs delivery), and it includes the SPIRE~350 and 500 \micron\ bands that we did not use in our SED fitting. We also provide the resolution matched, background subtracted maps for all galaxies for which we perform SED fitting.  These data products can be straightforwardly ingested into other SED fitting procedures which make different assumptions about the dust and radiation field properties.

Until the next far-infrared space telescope is launched, the set of nearby galaxies observed by \herschel, that we compile here, will remain a uniquely important resource for studying interstellar dust in galaxies.

\begin{acknowledgements}
We thank the referee for their careful reading of the paper and suggestions that improved its clarity.
The work of JC, KS, IC, AKL is supported by NASA ADAP grants NNX16AF48G and NNX17AF39G and National Science Foundation grant No.~1615728. JC acknowledges support from ERC starting grant \#851622 DustOrigin, and funding from the Belgian Science Policy Office (BELSPO) through the PRODEX project ``JWST/MIRI Science exploitation'' (C4000142239). IC thanks the support from the National Science and Technology Council grant 111-2112-M-001-038-MY3, and the Academia Sinica investigator award AS-IA-109-M02 (PI: Hiroyuki Hirashita).
EWK acknowledges support from the Smithsonian Institution as a Submillimeter Array (SMA) Fellow and the Natural Sciences and Engineering Research Council of Canada.

We are very grateful to the staff of IPAC for their support in hosting the large dataset presented in this paper.
We acknowledge the usage of the HyperLeda database (\url{http://leda.univ-lyon1.fr}).
This research has made use of 
the SIMBAD database, operated at CDS, Strasbourg, France \citep[][]{Wenger2000}, 
TOPCAT, an interactive graphical viewer and editor for tabular data \citep{2005ASPC..347...29T},
pandas \citep{McKinney_2010, McKinney_2011},
APLpy, an open-source plotting package for Python hosted at \url{http://aplpy.github.com},
SciPy \citep{Virtanen_2020},
ds9, a tool for data visualization supported by the Chandra X-ray Science Center (CXC) and the High Energy Astrophysics Science Archive Center (HEASARC) with support from the JWST Mission office at the Space Telescope Science Institute for 3D visualization,
matplotlib, a Python library for publication quality graphics \citep{Hunter:2007},
Astropy, a community-developed core Python package for Astronomy \citep{2018AJ....156..123A, 2013A&A...558A..33A},
NumPy \citep{harris2020array}.
This publication makes use of data products from the Wide-field Infrared Survey Explorer\citep{2010AJ....140.1868W}, which is a joint project of the University of California, Los Angeles, and the Jet Propulsion Laboratory/California Institute of Technology, funded by the National Aeronautics and Space Administration. 
\end{acknowledgements}
\begin{acknowledgements}
SPIRE has been developed by a consortium of institutes led by Cardiff University (UK) and including Univ. Lethbridge (Canada); NAOC (China); CEA, LAM (France); IFSI, Univ. Padua (Italy); IAC (Spain); Stockholm Observatory (Sweden); Imperial College London, RAL, UCL-MSSL, UKATC, Univ. Sussex (UK); and Caltech, JPL, NHSC, Univ. Colorado (USA). This development has been supported by national funding agencies: CSA (Canada); NAOC (China); CEA, CNES, CNRS (France); ASI (Italy); MCINN (Spain); SNSB (Sweden); STFC, UKSA (UK); and NASA (USA).
PACS has been developed by a consortium of institutes led by MPE (Germany) and including UVIE (Austria); KUL, CSL, IMEC (Belgium); CEA, OAMP (France); MPIA (Germany); IFSI, OAP/AOT, OAA/CAISMI, LENS, SISSA (Italy); IAC (Spain). This development has been supported by the funding agencies BMVIT (Austria), ESA-PRODEX (Belgium), CEA/CNES (France), DLR (Germany), ASI (Italy), and CICT/MCT (Spain).
This research has made use of HIPE, a joint development by the Herschel Science Ground Segment Consortium, consisting of ESA, the NASA Herschel Science Center, and the HIFI, PACS and SPIRE consortia.
\end{acknowledgements}
\facility{WISE, Herschel}

\bibliography{main}
\bibliographystyle{aasjournal}





\appendix

\section{Data Delivery}
\label{AppDelivery}

\subsection{Catalog}
We provide a machine readable table containing the parameters describing each galaxy, the integrated photometry, and our integrated SED fitting results for each target.  These are described in Table~\ref{tab:sample} and in detail, this includes:  
galaxy name; 
PGC name; 
Right Ascension and Declination; \rtf value;  effective radius $r_{\rm e}$;
inclination; position angle, distance, integrated flux in bands from WISE~1 to SPIRE~250 and associated errors;
the type of fit available (resolved or integrated, where resolved means integrated is available as well);
and the SED fitting parameter results including \umin, \ubar, $\gamma$, \qpah, and \mdust, and associated errors.

\subsection{Resolved maps}
We provide four sets of FITS files related to the resolved analysis from this work:
\begin{itemize}
    \item {\tt [GalaxyName]\_[Band].fits}: the \textit{Herschel} images after \scanam reduction and the background subtraction at native resolution, in MJy~sr$^{-1}$. The header of these maps contain \scanam keywords, and include the observation IDs from the \textit{Herschel} archive used to create the final map;
    In the header of these maps, we add a few key elements that were used in the data processing:
    \vspace{-.2cm}
    \begin{itemize}
    \setlength\itemsep{-0.1cm}
        \item {\tt R25COEFF}: the $A$ coefficient used to scale \rtf to create the galaxy mask (Section~\ref{SecBkgRemoval});
        \item {\tt COEFF[1,2,3]}: the coefficients of the plane to remove a 2D-background;
        \item {\tt ADDGAL[x]}: the name(s) of any galaxy that was found in the cut-out and consequently masked following identical procedure as described in Section~\ref{SecBkgRemoval};
    \end{itemize}
    \item {\tt [GalaxyName]\_[Band]\_conv250.fits}: the background-subtracted, convolved to SPIRE~250 $18''$ resolution, regridded \herschel maps, in MJy~sr$^{-1}$. The headers of these maps contain the same extra keywords as the ones mentioned just above;
    \item {\tt [GalaxyName]\_DustParameters\_conv250.fits}: a multi-extensions file containing the realizations maps \citep{Gordon14} of the \citet{DL07} dust parameters, \umin, $\gamma$, \qpah, \sigd, and \omgs, as well as \ubar (not fitted, calculated), for pixels passing a 1$\sigma$ S/N cut in the data (Section~\ref{SecFittingProcedure}), and their associated ${\rm 16^{th}}$--${\rm 84^{th}}$ percentiles. The maps and errors are calculated from 100 realization maps for each galaxy;
    \item {\tt [GalaxyName]\_extra.fits}: a multi-extension file containing:
    \begin{itemize}
    \vspace{-.2cm}
    \setlength\itemsep{-0.1cm}
        \item the galaxy mask, with the used {\tt R25COEFF} coefficient;
        \item masks of the pixels passing the 1-, 2-, and 3$\sigma$ S/N cuts;
        \item the final ``master-mask'' at the SPIRE~250 resolution, masking the galaxy, stars, and bright sources.
    \end{itemize}
\end{itemize}

\section{Dust Emission Parameters Corner Plot}\label{app:corner}
\begin{figure*}
    \centering
    \includegraphics[width=\textwidth, clip, trim={2cm 1cm 3cm 2.8cm}]{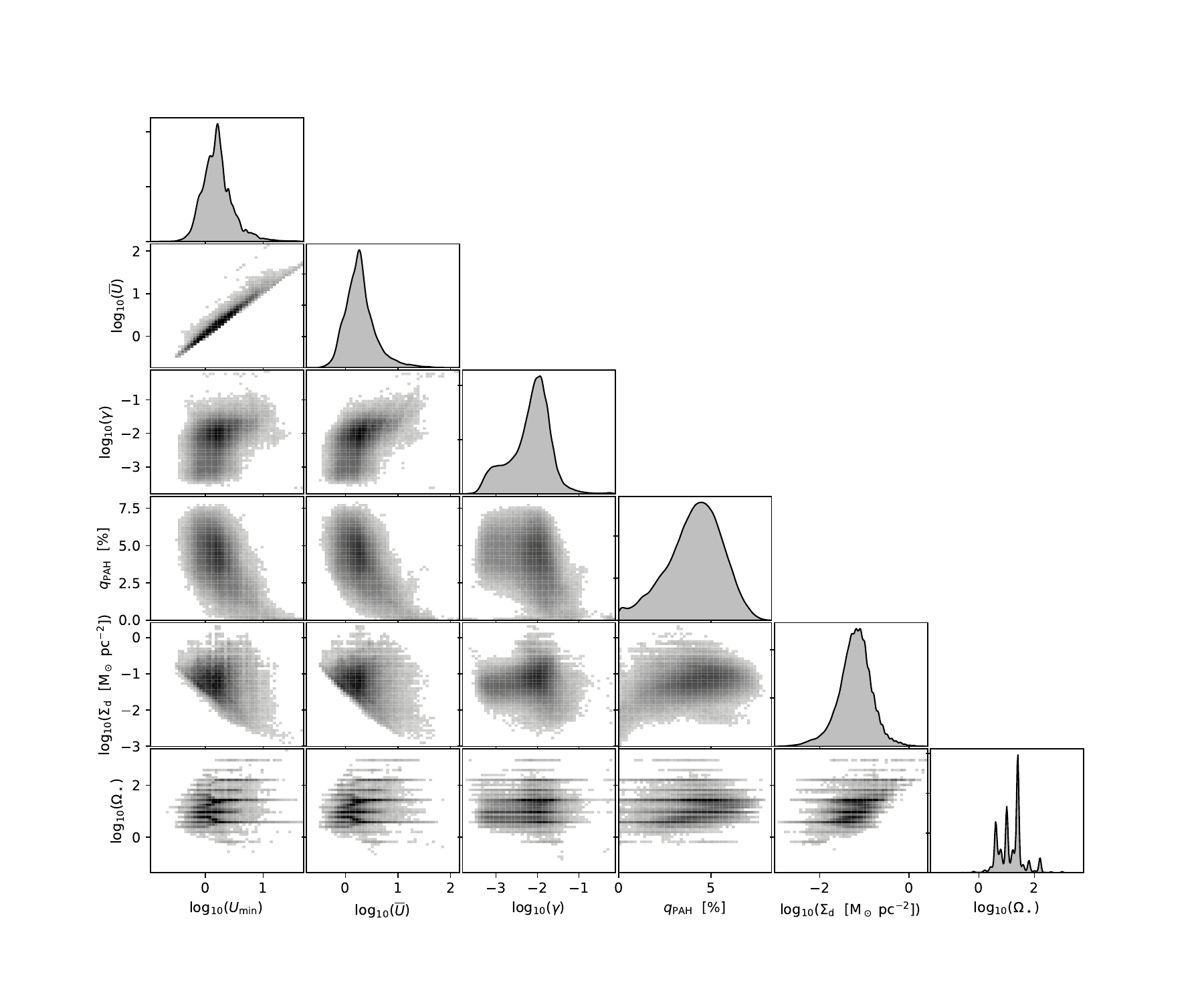}
    \caption{Distributions of the fitted dust and radiation field model parameters  for all resolved pixel in this analysis passing the cuts (Section~\ref{SecPixelsUsed}). The top plot of each column shows a marginalized 1d histogram of that parameter.  The other panels show two dimensional histograms of all pixels in the resolved analysis. Note that \ubar is not a fitted parameter, but is derived from the combination of \umin and $\gamma$.}
    \label{FigCornerPlot}
\end{figure*}

Figure~\ref{FigCornerPlot} shows the 2D and 1D marginalized histograms of the fitted dust and radiation field parameters and their relationships with each other.  
The clear relation between \umin and \ubar is expected due to the definition of \ubar, see Equation~\ref{EquUbar}. The scatter in the \ubar--\umin plot is correlated with $\gamma$. 
There is a degree of expected degeneracy between the model parameters \citep[extensively investigated in][]{Galliano2021}, which is visible in the Figure:
the far-IR SED responds strongly to a change in \umin, shifting the IR peak to shorter wavelengths, and in \sigd, scaling the SED vertically. This is seen in the anti-correlation between these two parameters, and is similar to the $\beta-T_{\rm d}$ relation investigated in the literature \citep[][]{Kelly2012, Galliano2021}.
Both these parameters, and the stellar light scaling factor, $\Omega_\star$, also affect the mid-IR SED, and influence the value of \qpah.

The diagonal cut visible in the \umin--\sigd and \ubar--\sigd panels is due to selection effects. The typical sensitivity of the \herschel maps can be translated into a combination of \ubar and \sigd.  For low \ubar, the only detectable dust emission will correspond to high \sigd values.  At high \ubar, even small \sigd can potentially be detected.  This results in the diagonal cut clearly seen in the \ubar-\sigd and \umin-\sigd panels.

The $\Omega_\star$ parameter scales a 5\,000~K blackbody to account for starlight, primarily affecting the shortest mid-IR wavelengths. For the purposes of our fitting, we chose to use a coarse grid in $\Omega_\star$ since it was not critical to our analysis. The coarseness of the $\Omega_\star$ grid allowed us to use finer sampling of the other SED parameters.  The visible stripes in the bottom row of panels are due to this coarse sampling.

\section{Effect of the wavelength coverage}
\label{sec:appendix500vs250}
As mentioned in the text, we only use data up to SPIRE~250, as a compromise between spectral coverage and resolution. In the range of temperature (radiation field) covered in this sample, the infrared peak will fall short of 250~$\mu$m and using SPIRE~250 only is sufficient to measure the far-IR peak. 
Several works have shown the effect of wavelength coverage and resolution on the recovery of dust parameters \citep[e.g.,][]{Aniano12}. The effect of spatial resolution was also investigated by \citet{Galliano2011}.

\begin{figure*}
    \centering
    \includegraphics[width=\linewidth]{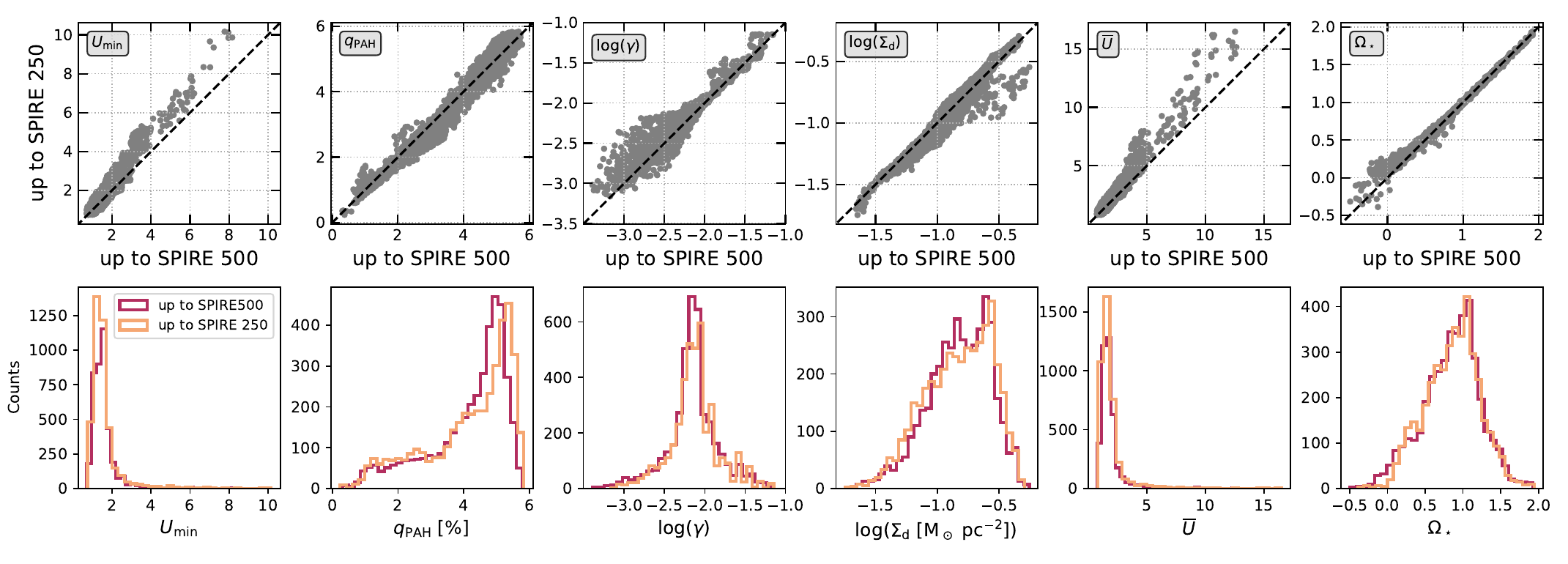}
    \caption{Distributions of the dust parameter values for M101 using the data from \citet{Chastenet21}. These values are extracted after fitting mid- to far-IR data at the SPIRE~500 resolution ($\sim 36''$) up to 500~$\mu$m, or up to 250~$\mu$m only.
    \textit{Top row:} values from a fit using all data up to 500~$\mu$m (x-axis) vs values from a fit using data up to 250~$\mu$m only (y-axis). The black-dashed line shows the 1:1 relation. Only the radiation field is noticeably affected by wavelength coverage, and the difference becomes large only at high \umin values.
    \textit{Bottom row:} histograms for the dust parameters.}
    \label{Fig500vs250}
\end{figure*}

As a test of the accuracy of fitting up to SPIRE 250, Figure~\ref{Fig500vs250} shows the distributions of the dust parameters for two fits at the same resolution of SPIRE~500, $\sim 36''$, using different spectral coverage: up to 500~$\mu$m and up to 250~$\mu$m. Data and results are those from \citet{Chastenet21}, covering the mid- to far-IR emission of M101 (NGC~5457).
The top row shows the pixel-by-pixel values for the two different fits. Only the radiation field values show a clear deviation from the 1:1 line, and it becomes significant (slope of $\sim1.25$) only at high \umin values. 
The bottom rows show that the distributions of values in both cases are virtually identical.

\end{document}

%% file: tab_delivery.tex
\renewcommand{\arraystretch}{1.1}
\begin{deluxetable*}{lll}
\caption{}
\label{tab:sample}
    \centering
    \tablehead{\multicolumn{1}{l}{Parameter} & \multicolumn{1}{l}{Unit} & \multicolumn{1}{l}{Description}}
\startdata
\hline
    {\tt GALNAME} & -- & Galaxy ``common'' identifier \\
    {\tt PGCNAME} & -- & Galaxy PGC identifier, from \galbase \\
    {\tt RA\_DEG} & \degree & Right ascension, from \galbase \\
    {\tt DEC\_DEG} & \degree & Declination, from \galbase \\
    {\tt POSANG\_DEG} & \degree & Position angle, from \galbase; if none, set to 0 during processing \\
    {\tt INCL\_DEG} & \degree & Inclination, from \galbase; if none, set to 0 during processing \\
    {\tt DIST\_MPC} & Mpc & Distance, from \galbase \\
    {\tt R25\_DEG} & \degree &  Optical radius, from \galbase \\
    {\tt REFF\_DEG} & \degree & Effective radius, from Sun et al. (in prep) \\
    {\tt FIT\_TYPE} & -- & Type of fit: {\tt I} indicates integrated, and {\tt R} resolved, which implies integrated fit available as well. \\
    {\tt (e)WISE1} & Jy & Integrated flux within \rtf in band WISE~1 \\
    {\tt (e)WISE2} & Jy & Integrated flux within \rtf in band WISE~2 \\
    {\tt (e)WISE3} & Jy & Integrated flux within \rtf in band WISE~3 \\
    {\tt (e)WISE4} & Jy & Integrated flux within \rtf in band WISE~4 \\
    {\tt (e)PACS70} & Jy & Integrated flux within \rtf in band PACS~70 \\
    {\tt (e)PACS100} & Jy & Integrated flux within \rtf in band PACS~100 \\
    {\tt (e)PACS160} & Jy & Integrated flux within \rtf in band PACS~160 \\
    {\tt (e)SPIRE250} & Jy & Integrated flux within \rtf in band SPIRE~250 \\
    {\tt (e)UMIN} & -- & Minimum radiation field from the integrated fit \\
    {\tt (e)GAMMA} & -- & Fraction of dust heated by a power-law combination of radiation field from the integrated fit \\
    {\tt (e)UBAR} & -- & Average radiation field, calculated from \umin and $\gamma$ \\
    {\tt (e)QPAH} & \% & Fraction of dust mass in the form of PAHs from the integrated fit \\
    {\tt (e)MDUST} & \msolpcsq & Total dust mass from the integrated fit \\   
    \hline
\enddata
\tablenotetext{}{The (e) denote the associated error column.}
\end{deluxetable*}{}

%% file: tab_integratedfits.tex
\renewcommand{\arraystretch}{1.3}
\begin{deluxetable*}{c|ccccc|cc}
    \centering
    \caption{Coefficients to the integrated galaxy fits, Y = $a$X + $b$, presented in Figures~\ref{FigIntStarU} and \ref{FigIntStarDust}.}\label{TabCoeffsFitsInt}
    \tablehead{& \multicolumn{5}{c}{Binned medians} & \multicolumn{2}{c}{All points}}
    \startdata
    Y = Dust parameter & $a \pm \epsilon_{a}^{\rm fit} \pm \epsilon_{a}^{\rm bin}$ & $b \pm \epsilon_{b}^{\rm fit} \pm \epsilon_{b}^{\rm bin}$ & Pearson's $\rho$ & RMSE & p-value & $a'$ & p-value$'$ \\
    \hline
      & \multicolumn{7}{c}{X = log$_{10}$(SFR)} \\ 
    \hline
    log$_{10}(U_{\rm min})$ & $0.21 \pm 0.04 \pm 0.03$  & $0.4 \pm 0.04 \pm 0.03$  & 0.88 & 0.437 & 0.002 & 0.17 &  $< 0.001$  \\ 
    log$_{10}(\gamma)$ & $0.27 \pm 0.01 \pm 0.03$  & $-1.89 \pm 0.02 \pm 0.01$  & 0.99 & 0.543 &  $< 0.001$  & 0.37 &  $< 0.001$  \\ 
    log$_{10}(\overline{U})$ & $0.25 \pm 0.04 \pm 0.02$  & $0.53 \pm 0.05 \pm 0.02$  & 0.88 & 0.444 & 0.002 & 0.25 &  $< 0.001$  \\ 
    $q_{\rm PAH}~~[\%]$ & $-0.72 \pm 0.29 \pm 0.24$  & $2.42 \pm 0.36 \pm 0.22$  & -0.28 & 1.898 & 0.5 & -0.03 & 0.72 \\ 
    log$_{10}({\rm M{_d}~~[M_\odot]}$) & $0.81 \pm 0.06 \pm 0.04$  & $7.25 \pm 0.06 \pm 0.02$  & 0.98 & 0.551 &  $< 0.001$  & 0.81 &  $< 0.001$  \\ 
    \hline
      & \multicolumn{7}{c}{X = log$_{10}$(M$_{\star}$)} \\ 
    \hline
    log$_{10}(U_{\rm min})$ & $-0.13 \pm 0.06 \pm 0.03$  & $1.56 \pm 0.57 \pm 0.32$  & -0.62 & 0.46 & $< 0.001$ & -0.02 & 0.58 \\ 
    log$_{10}(\gamma)$ & $0.12 \pm 0.03 \pm 0.04$  & $-3.08 \pm 0.32 \pm 0.45$  & 0.73 & 0.581 & $< 0.001$ & 0.26 &  $< 0.001$  \\ 
    log$_{10}(\overline{U})$ & $-0.06 \pm 0.05 \pm 0.03$  & $0.97 \pm 0.51 \pm 0.33$  & -0.4 & 0.492 & 0.2 & 0.05 & 0.1 \\ 
    $q_{\rm PAH}~~[\%]$ & $1.19 \pm 0.26 \pm 0.07$  & $-8.46 \pm 2.48 \pm 0.71$  & 0.88 & 1.644 &  $< 0.001$  & 0.77 &  $< 0.001$  \\ 
    log$_{10}({\rm M{_d}~~[M_\odot]}$) & $1.09 \pm 0.05 \pm 0.04$  & $-3.52 \pm 0.53 \pm 0.38$  & 0.99 & 0.466 &  $< 0.001$  & 1.0 &  $< 0.001$  \\ 
    \hline
      & \multicolumn{7}{c}{X = log$_{10}$(sSFR)} \\ 
    \hline
    log$_{10}(U_{\rm min})$ & $0.54 \pm 0.05 \pm 0.03$  & $5.74 \pm 0.54 \pm 0.28$  & 0.95 & 0.394 &  $< 0.001$  & 0.06 &  $< 0.001$  \\ 
    log$_{10}(\gamma)$ & $0.47 \pm 0.03 \pm 0.03$  & $2.77 \pm 0.35 \pm 0.3$  & 0.97 & 0.563 &  $< 0.001$  & 0.1 &  $< 0.001$  \\ 
    log$_{10}(\overline{U})$ & $0.7 \pm 0.06 \pm 0.06$  & $7.51 \pm 0.58 \pm 0.64$  & 0.93 & 0.394 &  $< 0.001$  & 0.09 &  $< 0.001$  \\ 
    $q_{\rm PAH}~~[\%]$ & $-2.58 \pm 0.23 \pm 0.14$  & $-22.86 \pm 2.28 \pm 1.4$  & -0.92 & 1.541 &  $< 0.001$  & -0.32 &  $< 0.001$  \\ 
    log$_{10}({\rm M{_d}~~[M_\odot]}$) & $0.08 \pm 0.1 \pm 0.04$  & $7.94 \pm 0.97 \pm 0.42$  & 0.18 & 0.836 & 0.5 & -0.11 & 0.001 \\ 
    \hline
      & \multicolumn{7}{c}{X = $\Delta$MS} \\ 
    \hline
    log$_{10}(U_{\rm min})$ & $0.44 \pm 0.05 \pm 0.05$  & $0.3 \pm 0.03 \pm 0.02$  & 0.93 & 0.372 &  $< 0.001$  & 0.46 &  $< 0.001$  \\ 
    log$_{10}(\gamma)$ & $0.51 \pm 0.04 \pm 0.04$  & $-1.95 \pm 0.02 \pm 0.01$  & 0.96 & 0.522 &  $< 0.001$  & 0.6 &  $< 0.001$  \\ 
    log$_{10}(\overline{U})$ & $0.57 \pm 0.06 \pm 0.06$  & $0.41 \pm 0.04 \pm 0.03$  & 0.94 & 0.352 &  $< 0.001$  & 0.6 &  $< 0.001$  \\ 
    $q_{\rm PAH}~~[\%]$ & $-1.89 \pm 0.26 \pm 0.28$  & $3.2 \pm 0.16 \pm 0.08$  & -0.89 & 1.578 &  $< 0.001$  & -1.25 &  $< 0.001$  \\ 
    log$_{10}({\rm M{_d}~~[M_\odot]}$) & $0.67 \pm 0.08 \pm 0.05$  & $7.26 \pm 0.05 \pm 0.04$  & 0.92 & 0.752 &  $< 0.001$  & 0.59 &  $< 0.001$  \\ 
    \enddata
    \tablecomments{The quoted ($a$, $b$) values are for the displayed number of bins, which are running medians of the whole sample; the first uncertainty, $\epsilon_x^{\rm fit}$, is the statistical error on these linear fits; the second uncertainty, $\epsilon_x^{\rm bin}$, is the standard deviation on coefficients ($a$, $b$) when fitting binned data with a varying number of bins.  Only the filled medians are considered for the fit, and chosen to have at least 10 points within the bin.
    The Pearson's $\rho$ coefficient is that of the fits shown. The root mean square error (RMSE) is calculated on all dark-grey points.
    The last two columns show $a'$ and its p-value, the slope derived by fitting \textit{all points}, without binning, and give an additional sense of the uncertainty on the fit.}
\end{deluxetable*}

%% file: tab_resolvedfits.tex
\renewcommand{\arraystretch}{1.3}
\begin{deluxetable*}{c|ccccc|cc}
    \centering
    \caption{Coefficients to the resolved fits, Y = $a$X + $b$, presented in Figures~\ref{FigStarU} and \ref{FigStarDust}.}\label{TabCoeffsFitsResolved}
    \tablehead{& \multicolumn{5}{c}{Binned medians} & \multicolumn{2}{c}{All points}}
    \startdata
    Y = Dust parameter & $a \pm \epsilon_{a}^{\rm fit} \pm \epsilon_{a}^{\rm bin}$ & $b \pm \epsilon_{b}^{\rm fit} \pm \epsilon_{b}^{\rm bin}$ & Pearson's $\rho$ & RMSE & p-value & $a'$ & p-value$'$ \\
    \hline
    \hline
       & \multicolumn{7}{c}{X = log$_{10}(\Sigma_{\rm SFR})$} \\ 
    log$_{10}(U_{\rm min})$ & $0.42 \pm 0.01 \pm 0.01$  & $1.08 \pm 0.02 \pm 0.02$  & 0.99 & 0.235 &  $< 0.001$  & 0.4 &  $< 0.001$  \\ 
    log$_{10}(\gamma)$ & $0.39 \pm 0.02 \pm 0.01$  & $-1.25 \pm 0.04 \pm 0.02$  & 0.98 & 0.442 &  $< 0.001$  & 0.53 &  $< 0.001$  \\ 
    log$_{10}(\overline{U})$ & $0.52 \pm 0.02 \pm 0.01$  & $1.36 \pm 0.03 \pm 0.01$  & 0.99 & 0.254 &  $< 0.001$  & 0.48 &  $< 0.001$  \\ 
    $q_{\rm PAH}~~[\%]$ & $-1.84 \pm 0.13 \pm 0.06$  & $0.43 \pm 0.21 \pm 0.08$  & -0.96 & 1.498 &  $< 0.001$  & -1.17 &  $< 0.001$  \\ 
    log$_{10}({\rm \Sigma_{d}~~[M_\odot~pc^{-2}]}$) & $0.49 \pm 0.02 \pm 0.01$  & $-0.09 \pm 0.03 \pm 0.02$  & 0.99 & 0.309 &  $< 0.001$  & 0.53 &  $< 0.001$  \\ 
    \hline
      & \multicolumn{7}{c}{X = log$_{10}$($\Sigma_{\star}$)} \\ 
    log$_{10}(U_{\rm min})$ & $0.21 \pm 0.01 \pm 0.01$  & $-0.21 \pm 0.03 \pm 0.03$  & 0.98 & 0.264 &  $< 0.001$  & 0.22 &  $< 0.001$  \\ 
    log$_{10}(\gamma)$ & $-0.12 \pm 0.01 \pm 0.01$  & $-1.82 \pm 0.03 \pm 0.03$  & -0.95 & 0.512 &  $< 0.001$  & -0.08 &  $< 0.001$  \\ 
    log$_{10}(\overline{U})$ & $0.21 \pm 0.01 \pm 0.01$  & $-0.16 \pm 0.03 \pm 0.03$  & 0.98 & 0.291 &  $< 0.001$  & 0.21 &  $< 0.001$  \\ 
    $q_{\rm PAH}~~[\%]$ & $-0.7 \pm 0.09 \pm 0.09$  & $6.04 \pm 0.26 \pm 0.22$  & -0.92 & 1.657 &  $< 0.001$  & -0.64 &  $< 0.001$  \\ 
    log$_{10}({\rm \Sigma_{d}~~[M_\odot~pc^{-2}]}$) & $0.29 \pm 0.03 \pm 0.04$  & $-1.63 \pm 0.07 \pm 0.08$  & 0.9 & 0.337 &  $< 0.001$  & 0.28 &  $< 0.001$  \\ 
    \hline
      & \multicolumn{7}{c}{X = log$_{10}$(sSFR$_{\rm R})$} \\ 
    log$_{10}(U_{\rm min})$ & $0.47 \pm 0.06 \pm 0.03$  & $4.87 \pm 0.59 \pm 0.27$  & 0.94 & 0.282 &  $< 0.001$  & 0.38 &  $< 0.001$  \\ 
    log$_{10}(\gamma)$ & $0.61 \pm 0.06 \pm 0.03$  & $3.99 \pm 0.56 \pm 0.3$  & 0.95 & 0.449 &  $< 0.001$  & 0.63 &  $< 0.001$  \\ 
    log$_{10}(\overline{U})$ & $0.66 \pm 0.09 \pm 0.05$  & $6.81 \pm 0.89 \pm 0.47$  & 0.93 & 0.327 &  $< 0.001$  & 0.47 &  $< 0.001$  \\ 
    $q_{\rm PAH}~~[\%]$ & $-2.71 \pm 0.33 \pm 0.21$  & $-22.9 \pm 3.24 \pm 2.13$  & -0.89 & 1.446 &  $< 0.001$  & -2.57 &  $< 0.001$  \\ 
    log$_{10}({\rm \Sigma_{d}~~[M_\odot~pc^{-2}]}$) & $-0.15 \pm 0.04 \pm 0.06$  & $-2.71 \pm 0.35 \pm 0.54$  & -0.72 & 0.377 & 0.01 & -0.22 &  $< 0.001$  \\ 
    \hline
      & \multicolumn{7}{c}{X = $\Delta$MS$_{\rm R}$} \\ 
    log$_{10}(U_{\rm min})$ & $0.54 \pm 0.02 \pm 0.01$  & $0.12 \pm 0.01 \pm 0.01$  & 0.99 & 0.239 &  $< 0.001$  & 0.5 &  $< 0.001$  \\ 
    log$_{10}(\gamma)$ & $0.61 \pm 0.03 \pm 0.02$  & $-2.21 \pm 0.02 \pm 0.01$  & 0.98 & 0.417 &  $< 0.001$  & 0.75 &  $< 0.001$  \\ 
    log$_{10}(\overline{U})$ & $0.69 \pm 0.03 \pm 0.01$  & $0.17 \pm 0.02 \pm 0.01$  & 0.99 & 0.258 &  $< 0.001$  & 0.61 &  $< 0.001$  \\ 
    $q_{\rm PAH}~~[\%]$ & $-2.31 \pm 0.16 \pm 0.13$  & $4.41 \pm 0.13 \pm 0.06$  & -0.97 & 1.389 &  $< 0.001$  & -1.94 &  $< 0.001$  \\ 
    log$_{10}({\rm \Sigma_{d}~~[M_\odot~pc^{-2}]}$) & $0.38 \pm 0.02 \pm 0.01$  & $-1.21 \pm 0.01 \pm 0.0$  & 0.98 & 0.347 &  $< 0.001$  & 0.43 &  $< 0.001$  \\  
    \enddata
    \tablecomments{The quoted ($a$, $b$) values are for the displayed number of bins, which are running medians of the whole sample; the first uncertainty is the statistical error on these linear fits; the second uncertainty is the standard deviation on coefficients ($a$, $b$) when fitting binned data with a varying number of bins.  Only the filled medians are considered for the fit, and chosen to have at least 250 points within the bin and above the completeness thresholds defined in the text (selection bias due to the 3$\sigma$ dust emission fit).
    The Pearson's $\rho$ coefficient is that of the fits shown. The RMSE is calculated on all dark-grey points.
    The last column shows $a'$, the slope derived by fitting \textit{all points}, without binning, and give an additional sense of the uncertainty on the fit.}
\end{deluxetable*}

%% file: tab_planefits.tex
\renewcommand{\arraystretch}{1.3}
\begin{deluxetable*}{c|cccc}
    \centering
    \caption{Coefficients of Plane Fits: Y = a$\times \log_{10}(\Sigma_\star)$ + b$\times \log_{10}(\Sigma_{\rm SFR})$ + c. }\label{TabCoeffs2d}
    \tablehead{Y = Dust parameter & a & b & c & RMSE}
    \startdata
    \hline
    log$_{10}(\overline{U})$ & $ -0.22 \pm 0.02 $ & $ 0.67 \pm 0.02 $ & $ 2.14 \pm 0.06 $ & 0.200 \\ 
    $q_{\rm PAH}~~$[\%] & $ 1.25 \pm 0.09 $ & $ -2.44 \pm 0.09 $ & $ -3.68 \pm 0.29 $ & 1.196 \\ 
    log$_{10}({\rm \Sigma_{d}~~[M_\odot~pc^{-2}]}$) & $ 0.41 \pm 0.03 $ & $ 0.22 \pm 0.02 $ & $ -1.42 \pm 0.08 $ & 0.257 \\ 
    \hline
    \hline
    \enddata
\end{deluxetable*}